\documentclass[12pt,preprint]{aastex}
\usepackage[version=3]{mhchem}
\usepackage{lscape}
\usepackage{longtable}

\makeatletter
\newcounter{reaction}
\renewcommand\thereaction{C\,\arabic{reaction}}
\newcommand\reactiontag{\refstepcounter{reaction}\tag{\thereaction}}
\newcommand\reaction@[2][]{\begin{equation}\ce{#2}%
\ifx\@empty#1\@empty\else\label{#1}\fi%
\reactiontag\end{equation}}
\newcommand\reaction@nonumber[1]{\begin{equation*}\ce{#1}%
\end{equation*}}
\newcommand\reaction{\@ifstar{\reaction@nonumber}{\reaction@}}
\makeatother

\shorttitle{Photochemistry of Terrestrial Exoplanet Atmospheres}
\shortauthors{Hu et al.}

\begin{document}

\title{Photochemistry in Terrestrial Exoplanet Atmospheres I: Photochemistry Model and Benchmark Cases}

\author{Renyu Hu$^1$, Sara Seager$^{1,2}$, William Bains$^{1,3}$}
\affil{$^1$Department of Earth, Atmospheric and Planetary Sciences, Massachusetts Institute of Technology, Cambridge, MA 02139}
\affil{$^2$Department of Physics, Massachusetts Institute of Technology, Cambridge, MA 02139}
\affil{$^3$Rufus Scientific, Melbourn, Royston, Herts, United Kingdom}
\email{hury@mit.edu}

\begin{abstract}
We present a comprehensive photochemistry model for exploration of the chemical composition of terrestrial exoplanet atmospheres.
The photochemistry model is designed from the ground up to have the capacity to treat all types of terrestrial planet atmospheres, ranging from oxidizing through reducing, which makes the code suitable for applications for the wide range of anticipated terrestrial exoplanet compositions.
The one-dimensional chemical transport model treats up to 800 chemical reactions, photochemical processes, dry and wet deposition, surface emission and thermal escape of O, H, C, N and S bearing species, as well as formation and deposition of elemental sulfur and sulfuric acid aerosols.
We validate the model by computing the atmospheric composition of current Earth and Mars and find agreement with observations of major trace gases in Earth's and Mars' atmospheres.
We simulate several plausible atmospheric scenarios of terrestrial exoplanets, and choose three benchmark cases for atmospheres from reducing to oxidizing.
The most interesting finding is that atomic hydrogen is always a more abundant reactive radical than the hydroxyl radical in anoxic atmospheres.
Whether atomic hydrogen is the most important removal path for a molecule of interest also depends on the relevant reaction rates.
We also find that volcanic carbon compounds (i.e., \ce{CH4} and \ce{CO2}) are chemically long-lived and tend to be well mixed in both reducing and oxidizing atmospheres, and their dry deposition velocities to the surface control the atmospheric oxidation states.
Furthermore, we revisit whether photochemically produced oxygen can cause false positives for detecting oxygenic photosynthesis, and find that in 1-bar \ce{CO2}-rich atmospheres oxygen and ozone may build up to levels that have been previously considered unique signatures of life, if there is no surface emission of reducing gases.
The atmospheric scenarios present in this paper can serve as the benchmark atmospheres for quickly assessing the lifetime of trace gases in reducing, weakly oxidizing, and highly oxidizing atmospheres on terrestrial exoplanets for the exploration of possible biosignature gases.
\end{abstract}


\section{Introduction}

The search and characterization of super Earths is flourishing. A large number of super Earths are being detected (e.g., Rivera et al. 2005; Udry et al. 2007; Forveille et al. 2009; L\'eger et al. 2009; Charbonneau et al. 2009; Mayor et al. 2009; Vogt et al. 2010; Rivera et al. 2010; Dawson \& Fabrycky 2010; Holman et al. 2010; Howard et al. 2011; Bonfils et al. 2011; Winn et al. 2011; Demory et al. 2011; Batalha et al. 2011; Lissauer et al., 2011; Borucki et al. 2011; Fressin et al 2012; Cochran et al. 2012). Attempts to observe super Earth atmospheres are growing (e.g., Batahla et al. 2011 for Kepler 10 b; Gillon et al. 2012 and Demory et al. 2012 for 55 Cnc e), and one super Earth/mini Neptune GJ 1214b is being observed in as much detail as possible (e.g., Bean et al. 2010; Croll et al. 2011; D\'esert et al. 2011; Berta et al. 2012). Although we distinguish between super Earths and mini Neptunes theoretically (super Earths are rocky planets with thin atmospheres whereas mini Neptunes are planets with massive gas envelopes), it is so far difficult to discriminate between the two categories observationally. Nonetheless, we focus on terrestrial planet atmospheres that include super Earths, since those planets hold the most interest on the path to finding and characterizing planets that might harbor life. In addition, the observational push to super Earth characterization has the potential to provide a handful of super Earth atmospheres to study in the coming years. In the more distant future, the community still holds hope that a direct-imaging space-based mission under the Terrestrial Planet Finder concept will enable planets to be observed in reflected light.

Super Earth atmospheres require understanding of photochemistry for their study. This is because the amounts of trace gases in the atmospheres depend critically on the component gas sources (surface emission) and sinks (chemical reactions initiated by UV photolysis, as well as surface deposition). This is in contrast to giant exoplanets, where there is no surface for emission, and where UV photolysis leads largely to perturbations on atmospheric composition, not to the definition of atmospheric composition. For super Earths and smaller terrestrial exoplanets, the key processes to consider in determining atmospheric composition are photolysis, kinetics of the reactions between atmospheric components, vertical diffusion of molecules, atmospheric escape, dry and wet deposition, and condensation and sedimentation of condensable species. An atmospheric spectrum may have strong features from spectroscopically active trace gases whose lifetime depends on photochemistry; some of these trace gases may be hallmarks for specific atmospheric scenarios. Observational signatures of non-equilibrium chemistry have been observed for several transiting hot Jupiters and 	"hot Neptunes" (e.g., Swain et al. 2008; Stevenson et al. 2010; Madhusudhan \& Seager 2011).

Theoretical studies of photochemistry on exoplanets have also been ongoing. One-dimensional photochemistry models have long
been used to understand the atmospheres of planets and moons in the Solar System (see Yung \& Demore 1999 and references therein),
and such models and their networks of chemical reactions have been the foundation for exoplanet models. Photochemistry models have been very successful in understanding the atmosphere of Earth (e.g., Kasting et al. 1985; Zahnle 1986; Seinfeld \& Pandis 2006 and references therein), Mars (e.g. Yung \& Demore 1999; Zahnle et al. 2008), Venus (e.g. Krasnopolsky \& Pollack 1994; Zhang et al. 2012), and Titan (e.g. Atreya et al. 2006). After the discoveries of transiting hot Jupiters, photochemistry models were adapted and applied to exoplanets. Liang et al. (2003) suggest that, driven by the photolysis of water, hot Jupiters have much higher H concentrations than Jupiter in their upper atmospheres. Zahnle et al. (2009a) and Line et al. (2010) investigated the hydrocarbon photochemistry in hot Jupiters' \ce{H2}-dominated atmospheres with a complete chemical network (as compared to the work of Liang et al. (2003) that was based on a simplified hydrocarbon chemistry scheme). Zahnle et al. (2009b) studied sulfur photochemistry in hot Jupiters' atmospheres, and suggested that \ce{H2S} photolysis becomes important at altitudes above the  $\sim$100 Pa pressure level. Moses et al. (2011) studied atmospheric chemistry of HD 209458b and HD 189733b, the most well-characterized hot Jupiters, with a complete chemical network involving C, H, O, N species and derived observational signatures of disequilibrium processes in their atmospheres. Kopparapu et al. (2012) presented a photochemistry model of a possibly carbon-rich hot Jupiter, WASP-12b, and suggested that \ce{C2H2} and \ce{HCN}, produced by photochemistry, may contribute significantly to the opacity of its atmosphere. More recently, Millier-Ricci Kempton et al. (2012) presented the first analysis of photochemistry in a \ce{H2}-dominated atmosphere of mini Neptune GJ 1214b.

A key challenge in developing a photochemistry model for terrestrial exoplanet atmospheres is handling the very wide range of plausible atmospheric compositions.  The broad possibilities for super Earth atmospheric composition arise from the ideas for atmospheric origin.  Super Earth atmospheres may originate from the capture of nebular gases, degassing during accretion, and/or degassing from subsequent tectonic activity (Elkins-Tanton \& Seager 2008; Schaefer \& Fegley 2010). In principle, the atmospheres of super-Earths and terrestrial exoplanets may be reducing (\ce{H2}-rich), oxidized (\ce{N2}-rich, or \ce{CO2}-rich), or oxidizing (\ce{O2}-rich). It is also possible that some super-Earth atmospheres are water rich (Rogers \& Seager 2010; Bean et al. 2010; Miller-Ricci Kempton et al. 2012). To date, photochemistry models are usually specific to certain types of atmospheres, because very different photochemical reactions dominate the chemistry of those atmospheres (for example in oxidized vs. reducing atmospheres).  In particular, photochemistry of hydrogen-dominated atmospheres on terrestrial planets is yet to be explored; but hydrogen-dominated atmospheres may still create habitable surface temperature through collision-induced absorption (Pierrehumbert \& Gaidos 2011; Wordsworth 2012).

A second significant challenge for a terrestrial exoplanet photochemistry model is the range of free parameters that describe key physical processes, namely vertical diffusion, fluxes of surface emission, and rates of dry and wet deposition. While the fundamental equations of continuity that govern the atmospheric composition are the same, these coefficients must be kept as free parameters in order to explore the range of atmospheric mixing and the range of plausible surface conditions, which yield the wide variety of possible exoplanets' atmospheres. The motivation of this work is to provide a photochemistry model that can be applied to study atmospheres of a wide range of redox states and that can handle the range of free parameters for physical processes in terrestrial exoplanet atmospheres.

In this paper, we present a photochemistry model developed from the ground up from basic chemical and physical principles and using both established and improved computer algorithms, which have the capacity of modeling very different terrestrial exoplanet atmospheres in a consistent way. The design of the photochemistry model aims at providing the maximum flexibility in treating atmospheres of very different compositions, especially oxidation states. Besides stellar parameters and planetary parameters (mass, radius, etc.), the photochemistry model accepts a series of input parameters including temperature structure of the atmosphere, dominant species and mean molecular mass of the atmosphere, coefficients of vertical diffusion, fluxes of surface emission and fluxes of dry and wet depositions. These input parameters are treated as free parameters so that the photochemistry model can simulate a wide range of atmospheric compositions.

In \S~2 we describe our photochemistry model in detail. In \S~3 we validate our photochemistry model by simulating the atmospheres of current Earth and Mars. In \S~4 we present three benchmark cases for exoplanets, representing reducing, weakly oxidizing, and highly oxidizing atmospheres. We discuss key parameters that determine atmospheric compositions in \S~4 and conclude in \S~5.

\section{A Photochemistry Model}

The purpose of the photochemical model is to compute the steady-state chemical composition of an exoplanetary atmosphere. The system can be described by a set of time-dependent continuity equations, one equation for each species at each altitude. Each equation describes: chemical production; chemical loss;  diffusion (contributing to production or loss); sedimentation (for aerosols only);  and specified gains and losses on the lower and upper boundaries of the atmosphere. The lower boundary has assigned geological or biological source fluxes and assigned deposition rates of species and the upper boundary has diffusion-limited atmospheric escape, relevant for light species.

With the assumption of a one-dimensional plane-parallel atmosphere, the unknowns to be computed in the model are the number densities of each species at each altitude. Starting from an initial state, the system is evolved to the steady state in which production and loss are balanced for each species at each altitude. Because the removal timescales of different species are very different, the inverse Euler method is employed for the numerical time stepping.  We allow the time step to be adjusted according to how fast the atmospheric components change (i.e., the variation timescale), and determine if the solution converges to the steady-state solution by evaluating the variation timescale for each species at each altitude, as well as the global production and loss balance for each species. Once the steady-state solution is found, we use a separate code to compute the exoplanet atmosphere spectrum for thermal emission, reflected stellar radiation, and transmission of stellar radiation.

We developed the photochemistry model from the ground up, and tested and validated the model by reproducing the atmospheric composition of Earth and Mars, in terms of globally-averaged 1D vertical mixing ratio profiles. For one of the tests, we started with a temperature-pressure profile of Earth's atmosphere, and with a subset of Earth's composition at present-day values, specifically \ce{N2}, \ce{O2}, and \ce{CO2} with assigned mixing ratios. We included surface emission of major trace gases, assumed to this exercise to be emitted according to typical globally averaged emission rates, for \ce{CO}, \ce{CH4}, \ce{NH3}, \ce{N2O}, \ce{NO}, \ce{NO2}, \ce{SO2}, \ce{OCS}, \ce{H2S}, and \ce{H2SO4}. The code correctly predicts the amounts of these trace gases in Earth's troposphere, and the amounts of ozone in the atmosphere as well as the location of the ozone layer. The code also produces vertical profiles of active hydrogen species (\ce{OH} and \ce{HO2}) and active nitrogen species (\ce{NO}, \ce{NO2}, and \ce{HNO3}) in the stratosphere that are consistent with observations. As another test, we reproduce the chemical composition of the current Mars atmosphere, in agreement with measured composition (e.g., Krasnopolsky 2006) and previous 1-D photochemistry model results (Nair et al. 1994; Yung \& DeMore 1999; Zahnle et al. 2008). In particular, the code correctly illustrate the effect of \ce{HO_x} catalytic chemistry that stabilizes Mars' \ce{CO2} atmosphere. The results of these tests are described in more detail below (Section 3).

The photochemistry model is designed for exoplanet applications, and has features that are not needed for studies of individual Solar System objects but which are useful for study of potential exoplanet atmospheres. The most important feature that yields the capacity of treating reducing and oxidized atmospheres with the same code is the flexibility of choosing chemical species to be considered in photochemical equilibrium. The species that have lifetimes less than the numerical time step can be treated as being in photochemical equilibrium. Our photochemistry model features a ``burn-in" period in which all species' number densities are rigorously computed and then those species that satisfy the lifetime condition are treated in equilibrium so that the computation can be optimized for speed. We have coded the Jacobian matrix for the implicit Euler method analytically, which improves the numerical rigor of the code. In this way our photochemistry model can quickly find the steady-state solution of atmospheric composition starting from any sensible initial conditions for both reducing and oxidized atmospheres. We emphasize that the ability to compute atmospheric composition of different oxidation states and the elimination of the need of providing fine-tuned initial conditions are essential for exoplanet exploration, because the oxidation states are unknown for terrestrial exoplanets and there are no benchmark cases to provide common initial conditions.

Another important feature of our photochemistry model is that it treats a wide range of free parameters for terrestrial exoplanets. We do not hardwire any model parameters that meant to be free for exoplanet exploration, including those parameters for planets, such as stellar spectrum and surface gravity, those parameters for atmospheres, such as mean molecular mass, refractive indices, temperature profiles (which can also be self-consistently computed), and eddy diffusion coefficients, and those parameters that are specific to particular molecules, such as solubility, rainout rate, saturation vapor pressure, surface emission rate, and dry deposition velocity. The code thus has a clear structure that makes extensive parameter exploration possible.

The third important feature of our photochemistry model is that we offer the flexibility of choosing a subset of chemical species and chemical reactions for the computation. It is in particular important for exoplanet exploration to be able to isolate a chemical system from the complex network of atmospheric chemistry; and our photochemistry model responds to this need. We design our code to be able to include or exclude the effect of any species or reactions in the complex chemical network, in order to understand the fundamentals of atmospheric chemistry on terrestrial exoplanets. Also, we pay attention to the temperature range in which chemical kinetic rates and photochemical cross sections are valid. The temperature range of exoplanet atmospheres can be significantly larger than the Solar System planet atmospheres. We label the chemical reactions whose rates are measured only at low temperatures ($T<400$ K) and the chemical reactions whose rates are measured only at high temperatures ($T>1000$ K), and use them only at appropriate temperature ranges. In addition, we consider the temperature dependence of photochemical cross sections in the model. Finally, we include a basic aerosol formation scheme in the model, the computation of which only relies on the saturation pressure of relevant substances and a timescale of condensation and sublimation. The aerosol scheme is simple, but could easily be expanded to treat non-conventional aerosols in the atmosphere.

We now present the detailed formulation of the photochemistry model.

\subsection{Fundamental Equations}

The coupled 1-dimensional continuity-transport equation that governs the chemical composition is
\begin{equation}
\frac{\partial n}{\partial t}=P-nL-\frac{\partial \Phi}{\partial z} , \label{eq_continuity}
\end{equation}
where $n$ is the number density of a certain species (cm$^{-3}$), $z$ is the altitude, $P$ is the production rate of the species (cm$^{-3}$ s$^{-1}$), $L$ is the loss rate of the species (s$^{-1}$), and $\Phi$ is the upward vertical transport flux of the species (cm$^{-2}$ s$^{-1}$). The flux can be approximated by the eddy diffusion together with the molecular diffusion, viz.,
\begin{equation}
\Phi=-KN\frac{\partial f}{\partial z} - DN\frac{\partial f}{\partial z}
+Dn\bigg(\frac{1}{H_0}-\frac{1}{H}-\frac{\alpha_{\rm T}}{T}\frac{dT}{dz}\bigg) ,\label{eq_diffusion}
\end{equation}
where $K$ is the eddy diffusion coefficient (cm$^2$ s$^{-1}$), $D$ is the molecular diffusion coefficient (cm$^2$ s$^{-1}$), $N$ is the total number density of the atmosphere, $f\equiv n/N$ is the mixing ratio of the species, $H_0$ is the mean scale height, $H$ is the molecular scale height, $T$ is the temperature (K), and $\alpha_{\rm T}$ is the thermal diffusion factor (Banks \& Kockarts 1973; Levine 1985; Krasnopolsky 1993; Yung \& DeMore 1999; Bauer \& Lammer 2004; Zahnle et al. 2006). The first term of equation (\ref{eq_diffusion}) represents eddy diffusion, and the last two terms represent molecular diffusion. The vertical transport flux is written in terms of the derivatives of the mixing ratio rather than in terms of the number density because this form is simpler and more straightforward to be implemented in numerical schemes. We show that equation (\ref{eq_diffusion}) is equivalent to the standard vertical diffusion equation in Appendix \ref{A_VD}.

The molecular diffusion coefficient $D$ and the thermal diffusion factor $\alpha_{\rm T}$ can be determined from the gas kinetic theory, but the eddy diffusion coefficient is a more speculative parameter and must be estimated empirically. For molecular diffusion, we use the following expressions in cm$^2$ s$^{-1}$ for \ce{H} and \ce{H2} in a \ce{N2}-based atmosphere
\begin{eqnarray}
D(\ce{H}, \ce{N2}) & = & \frac{4.87 \times10^{17} \times T^{0.698}}{N} ,\nonumber\\
D(\ce{H2}, \ce{N2}) & = & \frac{2.80 \times10^{17} \times T^{0.740}}{N} ; \label{mdiff1}
\end{eqnarray}
and in a \ce{CO2}-based atmosphere
\begin{eqnarray}
D(\ce{H}, \ce{CO2}) & = & \frac{3.87 \times10^{17} \times T^{0.750}}{N} ,\nonumber\\
D(\ce{H2}, \ce{CO2}) & = & \frac{2.15 \times10^{17} \times T^{0.750}}{N} ;
\end{eqnarray}
and in a \ce{H2}-based atmosphere
\begin{equation}
D(\ce{H}, \ce{H2})  =  \frac{8.16 \times10^{17} \times T^{0.728}}{N} . \label{mdiff2}
\end{equation}
In equations (\ref{mdiff1}-\ref{mdiff2}), $T$ has a unit of K and $N$ has a unit of cm$^{-3}$. 
The functional form of equations (\ref{mdiff1}-\ref{mdiff2}) is derived from the gas kinetic theory and the coefficients are obtained by fitting to experimental data (Marreo \& Mason 1972; Banks \& Kockarts 1973). The expressions for molecular diffusion coefficients are valid for a wide range of temperatures, up to 2000 K, except for $D(\ce{H2}, \ce{CO2})$ that is only valid at low temperatures, i.e., $T<550$ K (Marreo \& Mason 1972). $D(\ce{H}, \ce{CO2})$ is assumed to be 1.8 times larger than $D(\ce{H2}, \ce{CO2})$ (Zahnle et al. 2008). The thermal diffusion factor $\alpha_{\rm T}$ is taken as a constant for \ce{H} and \ce{H2} as $\alpha_{\rm T} = -0.38$ (Banks \& Kockarts 1973). The negative sign of $\alpha_{\rm T}$ corresponds to the fact that the lightest molecules tend to diffuse to the warmest region.

The eddy diffusion coefficient is the major uncertainty in the 1-D photochemistry model. For Earth's atmosphere, the eddy diffusion coefficient can be derived from the number density profile of trace gases (e.g. Massie \& Hunten 1981). The eddy diffusion coefficient of Earth's atmosphere is characterized by the convective troposphere and the non-convective stratosphere, which is not necessarily applicable for the exoplanets. The eddy diffusion coefficient may be typically parameterized as
\begin{eqnarray}
&& K=K_T\quad\quad (z<z_T) ,\nonumber\\
&& K={\rm min}\bigg(K_H\ ,\quad K_T\bigg(\frac{N(z_T)}{N}\bigg)^{1/2}\bigg)\quad\quad (z>z_T) ,\label{eddy}
\end{eqnarray}
where $K_T$, $K_H$ and $z_T$ are independent parameters satisfying $K_H>K_T$. This formula is adapted from Yung \& DeMore (1999) with the requirement of continuity in $K$. Our code can either import eddy diffusion coefficient from a file or specify eddy diffusion coefficient according to equation (\ref{eddy}).

The goal of the photochemistry model is to obtain a steady-state solution of each species, or a set of $n(z)$, which makes the left hand side of the equation (\ref{eq_continuity}) vanish and satisfies the boundary conditions. Assuming $N_x$ species in the model and $N_l$ equally stratified layers of the atmosphere, we transform the continuity equation (\ref{eq_continuity}) into a discrete form as
\begin{equation}
\frac{\partial n_i}{\partial t}=P_i-n_iL_i-\frac{\Phi_{i+1/2}-\Phi_{i-1/2}}{\Delta z} ,\label{eq_con_disc}
\end{equation}
where the subscript $i$ denotes physical quantities in the $i$th layer, and the subscripts $i+1/2$ and $i-1/2$ mean that the flux is defined at the upper and lower boundary of each layer. It is physically correct to define the flux term at the boundary of each layer, which is also crucial for numerically preserving hydrostatic equilibrium for an atmospheric transport scheme that uses number density as independent variables. According to equation (\ref{eq_diffusion}),
\begin{eqnarray}
\Phi_{i+1/2} & = & -(K_{i+1/2}+D_{i+1/2}) N_{i+1/2} \frac{f_{i+1}-f_i}{\Delta z} \nonumber\\
& & + D_{i+1/2}\frac{N_{i+1/2}}{2}\bigg[\frac{(m_a-m)g}{k_{\rm B}T_{i+1/2}}-\frac{\alpha_{\rm T}}{T_{i+1/2}}\frac{T_{i+1}-T_{i}}{\Delta z}\bigg] (f_{i+1}+f_{i})  , \label{diffusion_disc}
\end{eqnarray}
where $m_a$ is the mean molecular mass of the atmosphere, $m$ is the molecular mass of the species, $g$ is the gravitational acceleration, and $k_{\rm B}$ is the Bolzmann constant. We have approximated $f_{i+1/2}$ by $(f_{i+1}+f_{i})/2$ in equation (\ref{eq_diffusion}).

The combination of equation (\ref{eq_con_disc}) and (\ref{diffusion_disc}) gives the following 2nd ordered centered discrete differential equation to be solved numerically,
\begin{eqnarray}
\frac{\partial n_i}{\partial t} & = & P_i-n_iL_i
+ \bigg(k_{i+1/2}\frac{N_{i+1/2}}{N_{i+1}}-d_{i+1/2}\frac{N_{i+1/2}}{N_{i+1}}\bigg)n_{i+1}  \nonumber\\
& & - \bigg(k_{i+1/2} \frac{N_{i+1/2}}{N_i} +d_{i+1/2}\frac{N_{i+1/2}}{N_i} + k_{i-1/2} \frac{N_{i-1/2}}{N_i} - d_{i-1/2}\frac{N_{i-1/2}}{N_i}\bigg)n_i  \nonumber\\
& & +  \bigg(k_{i-1/2} \frac{N_{i-1/2}}{N_{i-1}}+d_{i-1/2} \frac{N_{i-1/2}}{N_{i-1}}\bigg)n_{i-1}  ,\label{Numerical}
\end{eqnarray}
where
\begin{eqnarray}
k_{i+1/2}  & = & \frac{K_{i+1/2}+D_{i+1/2}}{\Delta z^2} ,\nonumber\\
d_{i+1/2} & = & \frac{D_{i+1/2}}{2\Delta z^2} \bigg[\frac{(m_a-m)g\Delta z}{k_{\rm B}T_{i+1/2}}-\frac{\alpha_{\rm T}}{T_{i+1/2}}(T_{i+1}-T_i)\bigg]  . \nonumber
\end{eqnarray}
In principle, $\partial n/\partial t=0$ is equivalent to a set of $N_xN_l$ nonlinear algebraic equations. The set of nonlinear algebraic equations may be solved numerically by Newton-Raphson methods (Press et al. 1992). However, in practice, we find that Newton-Raphson methods require an initial guess to be in the vicinity of the solution. Instead, we treat the problem as a time-stepping problem by evolving the system according to equation (\ref{Numerical}) to the steady state.

Implicit numerical methods are implemented to solve equation (\ref{Numerical}). Due to the orders-of-magnitude differences in chemical loss timescales, the system is numerically stiff. We use the inverse-Euler method for the time stepping, as in most previous photochemistry models (e.g., Kasting et al. 1985; Nair et al. 1994; Zahnle et al. 2008). For each time step, we need to invert a matrix of dimension $N_xN_l$. As seen in equation (\ref{Numerical}), the variation in any layer only depends on the number density in that layer and adjacent layers. As a result, the matrix is by nature block tridiagonal with a block dimension of $N_x$, which is solved efficiently by the Thomas algorithm (Conte \& DeBoor 1972). The time step is self-adjusted in a way that the code updates the time step after each iteration according to the variation timescale of the whole chemical-transport system, i.e., the minimum variation timescale of each species at each altitude. As the system converges, larger and larger time steps are chosen.

The criteria of convergence to the steady-state solution is that the variation timescale of each species at each altitude is larger than the diffusion timescale of whole atmosphere, and the fluxes of gain and loss balance out for all species. The gain fluxes include surface emission, chemical production and condensation; whereas the loss fluxes include chemical loss, dry deposition, wet deposition, atmospheric escape, and condensation. We require that the ratio between the net global flux and the column-integrated number density for all long-lived species is small compared to the diffusion timescale. This condition is very important in the determination of convergence. If any possible long-term trends of major species can be detected, we run the code for an extended time period to test the convergence rigorously.

The current photochemistry model can compute concentrations of 111 molecules or aerosols made of C, H, O, N, S elements. These species are \ce{CO2}, \ce{H2}, \ce{O2}, \ce{N2}, \ce{H2O}, \ce{O}, \ce{O(^1D)}, \ce{O3}, \ce{H}, \ce{OH}, \ce{HO2}, \ce{H2O2}, \ce{N}, \ce{NH}, \ce{NH2}, \ce{NH3}, \ce{N2H2}, \ce{N2H3}, \ce{N2H4}, \ce{N2O}, \ce{NO}, \ce{NO2}, \ce{NO3}, \ce{N2O5}, \ce{HNO}, \ce{HNO2}, \ce{HNO3}, \ce{HNO4}, \ce{C}, \ce{CO}, \ce{CH4}, \ce{CH}, \ce{CH2}, \ce{CH2^1}, \ce{CH3}, \ce{CH2O}, \ce{CHO}, \ce{CH3O}, \ce{CH3O2}, \ce{CHO2}, \ce{CH2O2}, \ce{CH4O}, \ce{CH4O2}, \ce{C2}, \ce{C2H}, \ce{C2H2}, \ce{C2H3}, \ce{C2H4}, \ce{C2H5}, \ce{C2H6}, \ce{C2HO}, \ce{C2H2O}, \ce{C2H3O}, \ce{C2H4O}, \ce{C2H5O}, \ce{HCN}, \ce{CN}, \ce{CNO}, \ce{HCNO}, \ce{S}, \ce{S2}, \ce{S3}, \ce{S4}, \ce{S8}, \ce{SO}, \ce{SO2}, \ce{SO2^1}, \ce{SO2^3}, \ce{SO3}, \ce{H2S}, \ce{HS}, \ce{HSO}, \ce{HSO2}, \ce{HSO3}, \ce{H2SO4}, \ce{OCS}, \ce{CS}, \ce{CH3S}, \ce{CH4S}, \ce{CH3NO2}, \ce{CH3ONO2}, \ce{CH5N}, \ce{C2H2N}, \ce{C2H5N}, \ce{C3H2}, \ce{C3H3}, \ce{CH3C2H}, \ce{CH2CCH2}, \ce{C3H5}, \ce{C3H6}, \ce{C3H7}, \ce{C3H8}, \ce{C4H}, \ce{C4H2}, \ce{C4H3}, \ce{C4H4}, \ce{C4H5}, 1-\ce{C4H6}, 1,2-\ce{C4H6}, 1,3-\ce{C4H6}, \ce{C4H8}, \ce{C4H9}, \ce{C4H10}, \ce{C6H}, \ce{C6H2}, \ce{C6H3}, \ce{C6H6}, \ce{C8H2}, \ce{H2SO4} aerosols, \ce{S8} aerosols and organic hazes.

An important feature of the photochemistry code is the flexibility to choose a subset of molecules and aerosols to study a particular problem. Since the computation time for each time step scales with $N_x^3$, decreasing the number of chemical species in the main loop greatly reduces the computation time. As is common practice in the effort to reduce the stiffness of the system and improve the numerical stability, ``fast" species with relatively short chemical loss timescales are computed directly from the diagnostic equation, namely
\begin{equation}
n=\frac{P}{L}  ,\label{fast}
\end{equation}
which implies that photochemical equilibrium can be achieved within each time step. The choice of fast species varies with the atmospheric composition and should be considered on a case-by-case basis. The choices of ``fast" species were usually hard-wired in previous photochemistry codes, which fundamentally limited their application to certain specific types of atmospheres (e.g., to Earth and to Mars). Our photochemistry code offers the flexibility to adjust the choice of species in photochemical equilibrium, which yields the capacity of treating atmospheres having very different oxidation states with the same code. We evaluate $P$ and $L$ in equation (\ref{fast}) at each time step using the number density of all species from previous time step and determine the number density of fast species according to equation (\ref{fast}) after the current time step linearly. We neglect nonlinearity due to multiple species in photochemical equilibrium; as a result, we do not allow strongly inter-dependent species to be concurrently considered as in photochemical equilibrium, for example for \ce{S2}, \ce{S3}, and \ce{S4}. Our approach yields stable convergence to the steady-state solution from virtually any initial test solutions. We always verify the mass balance of the steady-state solution.

\subsection{Chemical Kinetics}

We compiled a comprehensive list of chemical and photochemical reactions from the literature. The production and loss rates in equation (\ref{eq_con_disc}) are provided from all chemical and photochemical reactions that produce or consume the relevant molecule. In the generic model, we included 645 bimolecular reactions, 85 termolecular reactions, and 93 thermal dissociation reactions. We included the thermal dissociation reactions to make the photochemistry model potentially adaptable to simulate hot planets. We mainly used the updated reaction rates from the online NIST database (http://kinetics.nist.gov) and refer to the JPL publication (Sander et al. 2011) for the recommended rate when multiple measurements are presented in the NIST database. For termolecular reactions, we used the complete formula suggested by the JPL publication (Sander et al. 2011) when the low-pressure-limiting rate and the high-pressure-limiting rate are available. For all reaction rates we use the Arrhenius formula to account for the dependence on temperature when the activation energy data are available, otherwise we adopted the value of experimental measurements, which are usually under $\sim$298 K.  We performed careful comparisons of our reaction lists and corresponding reaction rates with those used by Nair et al. (1994), Pavlov \& Kasting (2002), and Zahnle et al. (2008). Most of our reaction rates are the same as those used by other codes. For a dozen reactions, we find updated reaction rates listed in either the JPL publication or the NIST database updated after the publication of the cited photochemistry models. Also, for the general purpose of our photochemistry model, we include the reactions that are very slow at low temperatures (i.e., 200 - 400 K) but may be become important at higher temperatures $>1000$ K. In addition, a number of reactions that lack laboratory-measured rates may be important for low-temperature (i.e., 200 - 400 K) applications. We have adopted the rates for unmeasured sulfur reactions from Turco et al. (1982), Kasting (1990), and Moses et al. (2002), and the rates for reactions of C$>$2 hydrocarbons from Yung \& DeMore (1999). All chemical reactions and their reaction rates are tabulated in Table \ref{ReactionRates}.

Chemical reaction rates may sensitively depend on temperature. The reported reaction rates are usually valid in certain specific temperature ranges. For example, reaction rates recommended by the JPL publication (Sander et al. 2011) are the reactions relevant to Earth's atmosphere, in general valid within the small range of 200 - 300 K. Hence it may be problematic to extend the rate expressions to high-temperature cases, such as hot planets. At the high-temperature end, the valid temperature range for some of the reaction rates measured in the combustion chemistry is a few thousand K. These reactions may be important for modeling hot Jupiters (e.g., Moses et al. 2011), but they should not be included in the computation of low-temperature cases, e.g., for Earth-like planets. We annotate the low-temperature reactions and the high-temperature reactions in Table \ref{ReactionRates}. Moreover, most chemical reactions involving free radicals have very small activation energy, and thus weak temperature dependencies. Usually the reactions rate of free radicals are measured at room temperature and can be used in photochemistry models under different temperatures (Sander et al. 2011). Common atmospheric free radicals are: \ce{O(^1D)}, \ce{OH}, \ce{H}, \ce{N}, \ce{NH2}, \ce{N2H3}, \ce{C}, \ce{CH}, \ce{CH2}, \ce{CH2^1}, \ce{CH3}, \ce{CHO}, \ce{CHO2}, \ce{CH3O}, \ce{CH3O2}, \ce{C2}, \ce{C2H}, \ce{C2H3}, \ce{C2H5}, \ce{C2HO}, \ce{C2H3O}, \ce{C2H5O}, \ce{CN}, \ce{CNO}, \ce{S}, \ce{S2}, \ce{SO3}, \ce{HS}, \ce{HSO}, \ce{HSO2}, \ce{HSO3}, \ce{CH3S} and \ce{CS}. For reactions with these molecules, if no activation energy is reported in the literature, we assume that the activation energy is negligible and their kinetic rates do not depend on temperature.

\subsection{Interaction with Radiation}

Molecules in the atmosphere absorb ultraviolet (UV) and visible light from the host star. If the absorbed photon carries enough energy, the molecule may be photodissociated to form free radicals. The photodissociation rate is proportional to the number density of photons with UV and visible wavelengths at each altitude. For direct stellar radiation, the optical depth $\tau$ includes the contribution from molecular absorption $\tau_a$, Rayleigh scattering $\tau_r$, and aerosol particle extinction $\tau_m$. For the multiple-scattered (diffusive) radiation, we use the $\delta$-Eddington 2-stream method implemented based on the formulation of Toon et al. (1989). The actinic flux of the diffusive radiation is $F_{\rm diff} = 2(F^++F^-)$ where $F^+$ and $F^-$ are the diffusive flux in the upward and downward direction. The photolysis flux at a certain altitude includes both the direct radiation and the diffusive radiation, i.e.,
\begin{equation}
F(\lambda\ , z)=F_0(\lambda)\exp{[-\tau(\lambda\ , z)/\mu_0]} + F_{\rm diff} ,
\end{equation}
where $F_0$ is the radiation flux at the top of the atmosphere where $\tau=0$, and $\mu_0$ is the angle of the path of sunlight. By default we assume the zenith angle of the star to be $57.3^{\circ}$ (see Appendix \ref{A_Zenith} for justification). The photodissociation rate is then
\begin{equation}
J(z)=\frac{1}{2}\int q(\lambda)\sigma_a(\lambda)L(\lambda\ , z)d\lambda ,
\end{equation}
where $\sigma_a$ is the absorption cross section, $q(\lambda)$ is the quantum yield that is defined as the ratio between the yield of certain photodissociation products and the number of photons absorbed, and $L(\lambda\ , z)$ is the actinic flux with units of quanta cm$^{-2}$ s$^{-1}$ nm$^{-1}$\footnote{The actinic flux is the quantity of light available to molecules at a particular point in the atmosphere and which, on absorption, drives photochemical processes in the atmosphere. It is calculated by integrating the spectral radiance over all directions of incidence of the light. The actinic flux is distinguished from the spectral irradiance in the way that it does not refer to any specific orientation because molecules are oriented randomly in the atmosphere.}. The 1/2 factor is included to account for diurnal variation (e.g. Zahnle et al. 2008), which is not used to model the dayside of a tidally lock planet. With the general-purpose model, we treat 70 photodissociation reactions. Different branches resulting from the photodissociation of one molecule are treated as different photodissociation reactions. The photodissociation reactions, the sources of data for cross sections and quantum yields, and the rates on the top of Earth's atmosphere are tabulated in Table \ref{Photolysis}.

For the UV and visible cross sections and the quantum yields, we use the recommended values from the JPL publication (Sander et al. 2011) when available. We also use the cross sections from the MPI-Mainz-UV-VIS Spectral Atlas of Gaseous Molecules\footnote{www.atmosphere.mpg.de/spectral-atlas-mainz} when the JPL recommended values are not available or incomplete. There are a number of molecules of atmospheric importance that lack UV and visible cross sections and quantum yields, and we have estimated photolysis rates for them in the following way. The photodissociation timescale of \ce{S2} has been measured to be $\sim250$ s at 1 AU from the Sun (DeAlmeida \& Singh 1986), from which we estimate the photolysis rate of \ce{S2}. For other molecules that have no UV or visible-wavelength cross sections available in the literature, we assume their photolysis rate to be the same as another molecule that has similar structure. For example, the photolysis rate of \ce{HNO} is assumed to be the same as \ce{HNO2} (Zahnle et al. 2008); the photolysis rate of \ce{HSO} is assumed to be the same as \ce{HO2} (Pavlov \& Kasting 2002); and the photolysis rate of \ce{N2H2} is assumed to be the same as \ce{N2H4}.

Temperature dependencies of photolysis cross sections and quantum yields are considered. Notably, at 200 K, compared to room temperature, \ce{N2O}, \ce{N2O5}, \ce{HNO3}, \ce{OCS}, \ce{CO2} have smaller UV cross sections, leading to photolysis rates more than 10\% lower; whereas \ce{NO3} has a larger UV cross section, leading to a photolysis rate more than 10\% higher. \ce{SO2} has complex band structures in its UV spectrum and the cross sections depend on temperature as well. We take into account any other temperature dependencies reported for a gas at temperatures outside 290 - 300K (see Table \ref{Photolysis} for notes on temperature dependencies). In most cases, cross sections are also measured at lower temperature such as 200 K, primarily for Earth investigations (Sander et al. 2011). We use linear interpolation to simulate cross sections between 200 and 300 K, and do not extrapolate beyond this temperature range. It is worthwhile noting that UV cross sections of almost all gases at temperatures significantly higher than room temperature are unknown, and temperature dependencies of cross sections are not considered in previous high-temperature photochemistry models of hot exoplanets (e.g., Moses et al. 2011; Miller-Ricci Kempton et al. 2012). This may be a plausible simplification for gases without significant band structure in UV, but for gases such as \ce{H2}, \ce{CO2} and \ce{SO2}, whose band structures are sensitive to temperature, extrapolation of cross sections to high temperature might induce significant errors. One needs to be cautious about the uncertainty of photolysis rate at temperatures much higher than room temperature.

Rayleigh scattering from atmosphere molecules introduces additional optical depth, particularly important for attenuation of UV radiation. The optical depth due to Rayleigh scattering is
\begin{equation}
\tau_r(\lambda\ , z)=\int_z^{\infty}N(z')\sigma_r(\lambda) dz' ,
\end{equation}
in which the Rayleigh scattering cross section is (e.g., Liou 2002)
\begin{equation}
\sigma_r(\lambda) = C_r \frac{8\pi^3(m_r(\lambda)^2-1)^2}{3\lambda^4N_s^2} ,
\end{equation}
where $m_r$ is the real part of the refractive index of the molecule, $C_r$ is a corrective factor to account for the anisotropy of the molecule, and $N_s$ is the number density at the standard condition (1 atm, 273.15 K).
The refractive index depends on the main constituent in the atmosphere. The refractive index of Earth's atmosphere is from Seinfeld \& Pandis (2006); the refractive index of \ce{H2} is from Dalgarno \& Williams (1962); the refractive index of \ce{N2} is from Cox (2000); the refractive index of \ce{CO2} is from Old et al. (1971); and refractive indices of \ce{CO}, and \ce{CH4} are given in Sneep \& Ubachs (2005). In principle the correction factor $C_r$ depends on the molecule and the wavelength; but $C_r$ is usually within a few percent with respect to the unity, except for $C_r\sim1.14$ for \ce{CO2} (Sneep \& Ubachs 2005). In the following we assume $C_r=1.061$, the value for Earth's atmosphere at $\sim200$ nm (Liou 2002).

For the spectrum of a solar-type star (G2V), we use the Air Mass Zero (AM0) reference spectrum produced by the American Society for Testing and Materials\footnote{http://rredc.nrel.gov/solar/spectra/am0/}. The AM0 spectrum covers a wavelength range from 119.5 nm to 10 microns. For the extreme-UV spectrum, we use the average quiet-Sun emission provided by Curdt et al. (2004).

\subsection{Treatment of Aerosols}

Microphysical processes involved in the formation of atmospheric aerosols are nucleation, condensational growth, and coagulation (Toon \& Farlow 1981; Seinfeld \& Pandis 2006). It is beyond the purpose of this work to simulate these microphysical processes in detail. For photochemically produced aerosols, the competition between coagulation and sedimentation mainly determines the particle size distribution. A complete treatment of the atmospheric aerosols involves solving the steady-state size distribution function (e.g. Seinfeld \& Pandis 2006).

In our photochemical model, we simplify the problem by assuming the particle radius to be a free parameter. In Earth's atmosphere, aerosols formed in the atmosphere often have a lognormal size distribution around 0.1 - 1 $\mu$m (Seinfeld \& Pandis 2006). We define a particle radius parameter, $r_{\rm p}$, to be the surface area average radius. This parameterization allows us to separate the complexity of aerosol formation from the photochemistry model, as well as to explore how the particle size affects the overall chemical composition.

We compute the production and loss rate of a molecule in the condensed phase (i.e., aerosols) based on its condensation timescale. When the molecule becomes supersaturated at a certain altitude, condensation can happen and aerosols form. The condensation/evaporation timescale is given by Hamill et al. (1977) and Toon \& Farlow (1981) as
\begin{equation}
\frac{1}{t_c}=\frac{m}{4\rho_{\rm p}}\bigg(\frac{8k_{\rm B}T}{\pi m}\bigg)^{1/2}\frac{n_{\rm g}-n_{\rm v}}{r_{\rm p}} , \label{Tcond}
\end{equation}
where $t_c$ is the condensation/evaporation timescale, $m$ is the mass of molecule, $\rho_{\rm p}$ is the particle density, $k_{\rm B}$ is the Boltzmann constant, $T$ is the atmospheric temperature, $n_{\rm g}$ is the number density of the corresponding gas, and $n_{\rm v}$ is the saturated vapor number density at the corresponding pressure. The formula (\ref{Tcond}) is suitable for both condensation and evaporation processes. When $n_{\rm g}>n_{\rm v}$, the gas phase is saturated, so the condensation happens and $t_c>0$. When $n_{\rm g}<n_{\rm v}$, the gas phase is unsaturated, so the evaporation is possible and $t_c<0$. The production or loss rate of the molecule in the condensed phase is
\begin{equation}
P=\frac{n_g}{t_c}, \quad\quad L=\frac{1}{t_c} .\label{Rcond}
\end{equation}

We include gravitational settling in the mass flux term of the continuity-transport equation (equation \ref{eq_continuity}) for aerosol particles in addition to eddy mixing. The additional gravitational downward flux of the aerosol particle is
\begin{equation}
\Phi_{\rm F} = -v_{\rm F} n_c  ,\label{FluxAER}
\end{equation}
where $v_{\rm F}$ is the settling velocity of the particle in the atmosphere. The settling velocity is reached when the gravitational force is balanced by the gas drag. For aerosols with diameter of order of 1 $\mu$m, the settling velocity is reached within $10^{-5}$ s in Earth's atmosphere (Seinfeld \& Pandis 2006). Therefore we assume the falling velocity to be the settling velocity. The settling velocity can be derived from the Stokes' law (Seinfeld \& Pandis, 2006) as
\begin{equation}
v_{\rm F} = \frac{2}{9}\frac{r_{\rm p}^2\rho_{\rm p}gC_{c}}{\mu} ,\label{VFall}
\end{equation}
where $g$ is the gravitational acceleration, $\mu$ is the viscosity of the atmosphere, and $C_c$ is the slip correction factor related to the mean free path ($\lambda$) of the atmosphere as
\begin{equation}
C_c = 1+\frac{\lambda}{r_{\rm p}} \bigg[1.257+0.4\exp\bigg(-\frac{1.1r_{\rm p}}{\lambda}\bigg)\bigg]  .
\end{equation}

The treatment of photochemically produced aerosols described above is applicable to any molecules that could reach saturation as a result of photochemical production. In Earth's atmosphere, the photochemical aerosols include sulfate aerosols (\ce{H2SO4}), sulfur aerosols (\ce{S8}), organic hazes, nitric acid aerosols (\ce{HNO3}), and hydrochloric acid aerosols (\ce{HCl}). For now we have implemented sulfate aerosols (\ce{H2SO4}) and sulfur aerosols (\ce{S8}). These aerosols and water vapor are the common condensable materials at habitable temperatures that commonly exist in planetary atmospheres (e.g., Kasting et al. 1989; Pavlov et al. 2000; Seinfeld \& Pandis 2006). The required data for including an aerosol species in the photochemistry model is the saturation vapor pressure. Saturation vapor pressure of \ce{H2SO4} is taken as recommended by Seinfeld \& Pandis (2006) for atmospheric modeling, with a validity range of 150 - 360 K. Saturation pressure of \ce{S8} is taken as the total sulfur saturation pressure against liquid sulfur at $T>392$ K and solid (monoclinic) sulfur at $T<392$ K tabulated and expressed by Lyons (2008). In addition to aerosols, we use equation (\ref{Tcond}-\ref{Rcond}) to compute the process of condensation of water vapor in the atmosphere. We do not consider evaporation of condensed water in the atmosphere because water droplets may grow by aggregation and rapidly precipitate out (Seinfeld \& Pandis 2006). Saturation pressure of water is taken as that against ice at temperature lower than 273.16 K (Murphy \& Koop 2005), and that against liquid water at temperature higher than 273.16 K (Seinfeld \& Pandis 2006).

We treat the optical effect of aerosols by considering both scattering and absorption. Assuming a homogeneous sphere, the Mie theory (see Van de Hulst, 1981 for a detail description) computes extinction cross section ($\sigma_{\rm ext}$), single scattering albedo ($w_{\rm s}$) and asymmetric factor ($g_{\rm asym}$), based on the following parameters: the refractive index of the material $(m_r+m_ii)$,  $r_{\rm p}$, and the wavelength. In this paper, we use the refractive index of \ce{S8} aerosols from Tian et al. (2010) for the UV and visible wavelengths and from Sasson et al. (1985) for infrared (IR) wavelengths. We use the refractive index of \ce{H2SO4} aerosols (assumed to be the same as 75\% sulfuric acid solution) from Palmer \& William (1975) for  UV to IR wavelengths, and Jones (1976) for far IR wavelengths.

\subsection{Boundary Conditions}

\label{A_BC}

The atmospheric chemical composition of a terrestrial exoplanet is ultimately determined by boundary conditions. The upper boundary conditions describe the atmospheric escape. The lower boundary conditions describe the surface emission, the deposition of molecules and aerosols to the surface, or the presence of a large surface reservoir of certain molecule (e.g. \ce{H2O}). The boundary conditions need to be properly provided to capture the physics of an exoplanet atmosphere.

The upper boundary conditions describe the atmospheric escape of an terrestrial exoplanet. The escape rates of exoplanet atmospheres are fairly uncertain depending on stellar soft X-ray and UV luminosity, exosphere chemistry, existence of magnitude fields, etc (e.g., Yelle et al. 2008; Tian 2009). We thus provide the following options of specifying escape rates in the photochemistry code:
\begin{enumerate}
\item[] Type 1: $\Phi_{N_l+1/2}=0$, or no escape;
\item[] Type 2: $\Phi_{N_l+1/2}=n_{N_l} V_{\rm lim}$, where $V_{\rm lim}$ is the diffusion-limited escape velocity;
\item[] Type 3: $\Phi_{N_l+1/2}$ is a assigned nonzero value.
\end{enumerate}
Here we use the same notation for flux as in equation (\ref{eq_con_disc}-\ref{diffusion_disc}), such that the upper boundary condition replaces the flux at the upper boundary of the layer $N_l$ in equation (\ref{diffusion_disc}).
For the type 2 upper boundary condition (atmospheric escape), the diffusion-limited velocity ($V_{\rm lim}$) is
\begin{equation}
V_{\rm lim}=D_{N_l+1/2}\bigg(\frac{1}{H_0}-\frac{1}{H}\bigg)  ,
\end{equation}
where $D_{N_l+1/2}$ is the molecular diffusion coefficient evaluated at the top of the atmosphere. The diffusion-limited flux is the highest escape flux of an atmosphere in hydrostatic equilibrium (Hunten 1974). For Mars, the Jeans escape of \ce{H2} reaches the diffusion-limited flux when the exobase temperature is above 400 K (Zahnle et al. 2008). We use the Type 2 upper boundary condition for the escape of \ce{H} and \ce{H2} in our model, and generally we assume no escape for all other species, i.e., the Type 1 upper boundary condition. The Type 3 upper boundary condition may be used when processes above the neutral atmosphere are important. For example, an influx of atomic \ce{N} can represent the photodissociation of \ce{N2} in the upper atmosphere. The Type 3 boundary condition may also be used when hydrodynamic escape has to be considered.

The lower boundary conditions describe the interaction between the atmosphere and the surface, which includes surface emission and surface deposition. The three types of lower boundary conditions are:
\begin{enumerate}
\item[] Type 1: $n_{1}$ is assigned;
\item[] Type 2: $\Phi_{1/2}=-n_1V_{\rm DEP}$ where $V_{\rm DEP}$ is molecule-specific dry deposition velocity;
\item[] Type 3: $\Phi_{1/2}$ is assigned.
\end{enumerate}
Again we use the same notation for flux as in equation (\ref{eq_con_disc}-\ref{diffusion_disc}), and $\Phi_{1/2}$ replaces the flux at the lower boundary of the layer 1 in equation (\ref{diffusion_disc}).

The Type 1 lower boundary condition presents a large reservoir at the surface and for this paper we use this condition for water vapor to simulate the effect of a surface with oceans. This approach is equivalent to setting the relative humidity at the surface to be a constant. Note that specifying the Type 1 lower boundary condition means decreasing the number of free variables; $n_1$ is fixed as the lower boundary condition and is no longer considered as a variable in the main computation loop.

The Type 2 lower boundary condition specifies the dry deposition velocity, which is a key parameter that determines the chemical composition of the atmosphere and is a major unknown. The deposition velocity depends on both the dynamical properties of the lower atmosphere and the chemistry of the planet's surface. With the number density at the bottom layer ($n_1$) computed from the photochemistry model, the interaction between the bottom layer of atmosphere and the surface consists of two steps: first, the molecular transport across a thin stagnant layer of air adjacent to the surface, called the {\it quasi-laminar sublayer} ; second, the uptake at the surface (Seinfeld \& Pandis 2006). A parameterization of dry deposition velocity involving these two steps is described in Appendix \ref{A_DD}.

Physically the dry deposition requires a sink at the surface; for a gas without effective surface sink the dry deposition velocity should be zero. In particular, the surface deposition of a number of gases, including \ce{CO}, \ce{H2}, \ce{CH4}, and \ce{NH3}, is primarily removed by microorganisms on Earth (e.g., Kharecha et al. 2005; Seinfeld \& Pandis 2006; Kasting, 2012, private communication), and the canonical values for their dry deposition velocities are not applicable for presumably abiotic planets. Take carbon monoxide for an example. If a planet has no ocean, there is no known reaction that can consume \ce{CO} at the surface, and therefore the dry deposition velocity of \ce{CO} for an abiotic desiccated planet should be zero (like Mars). If the surface has an ocean, the dissolved \ce{CO} may be naturally and slowly converted into acetates by \ce{OH} in sea water, or biologically converted into acetates at a much faster rate. The \ce{CO} deposition velocity has been estimated to be $1.2\times10^4$ cm s$^{-1}$ for the most efficient biological removal and $10^{-8}\sim10^{-9}$ cm s$^{-1}$ on an abiotic ocean planet (Kasting 1990; Kharecha et al. 2005). For another example, it is common to use a fairly large dry deposition velocity for \ce{SO2} ($\sim1$ cm s$^{-1}$) to study the sulfur cycle in Earth's marine atmosphere (e.g., Toon et al. 1987). In contrast, the dry deposition velocity is assumed to be zero, or reduced by an arbitrary factor of up to 1000, to mimic a putative Mars ocean that is believed to be saturated with dissolved \ce{SO2} and other sulfur species (e.g., Halevy et al. 2007; Tian et al. 2010). In general, the dry deposition velocity depends on a broad context of planetary geochemistry, notably on the surface mineralogy, the acidity of ocean, and the surface pressure and temperature. Therefore it is critical for exoplanet exploration to understand the interaction between atmospheric chemistry composition of terrestrial exoplanets and the surface deposition.

In addition to the dry deposition at the surface, we also include the wet deposition throughout the atmosphere as a removal process for soluble species. We use the parameterization of Giorgi \& Chameides (1985) as
\begin{equation}
k_R(z)=f_R\times\frac{n_{\ce{H2O}}(z) k_{\ce{H2O}}(z)} {55A_V[L\times10^{-9}+(H'RT(z))^{-1}]}  ,\label{Rainout}
\end{equation}
where $f_R$ is a reduction factor adjustable in the model, $k_{\ce{H2O}}$ is the precipitation rate taken to be $2\times10^{-6}$ s$^{-1}$, $A_V$ is the Avogadro's number, $L$ is the liquid water content taken to be 1 g m$^{-3}$ in the convective layer near the surface, and $H'$ is the effective Henry's Law constant measuring the solubility of the molecule in the unit of mol dm$^{-3}$ atm$^{-1}$. The effective Henry's law constant may differ from the standard Henry's law constant when taking into account dissociation in the aqueous phase (Giorgi \& Chameides 1985; Seinfeld \& Pandis 2006). In the model we use the effective Henry's Law constant published in Giorgi \& Chameides (1985) as well as the standard Henry's Law constants from the NIST Chemistry Webbook (http://webbook.nist.gov/chemistry/). Since the parameterization of equation (\ref{Rainout}) is primarily for modeling Earth's atmosphere, the specific reduction factor $f_R$ should be applied when it is reasonable to believe the hydrological cycle is reduced on an exoplanet (e.g., Tian et al. 2010).

The Type 3 lower boundary condition has assigned flux that represents the surface emission. Note that the dry deposition (Type 2) and assigned flux (Type 3) lower boundary conditions can be used at the same time, but assigned number density (Type 1) boundary condition over-rules other lower boundary conditions. For example, \ce{SO2} may be deposited to the surface at a rate proportional to the number density of the bottom layer, and also be emitted from the surface to the atmosphere at an assigned flux.

We finish this section with a definition of the so-called ``redox balance", the requirement of which arises when the fixed mixing ratio lower boundary condition (Type-1) is used in the photochemistry model. Following Kasting \& Brown (1998), Zahnle et al. (2006), and Segura et al. (2007), we define \ce{H2O}, \ce{N2}, \ce{CO2}, \ce{SO2} as redox neutral, and assign the redox number (${\cal R}$) of any H-O-C-N-S molecule as the number of hydrogen in excess, i.e.,
\begin{equation}
{\cal R}(\ce{H_{a_1}O_{a_2}N_{a_3}C_{a_4}S_{a_5}}) = a_1 - 2a_2 + 4a_4 + 4a_5. \label{redox}
\end{equation}
Note that our definition of the redox number differs from Zahnle et al. (2006) and Segura et al. (2007) in the way that we count the number of \ce{H} and they count the number of \ce{H2}. With our definition, for example, the redox number of \ce{H2S} is 6, and the redox number of \ce{H2SO4} is -2. The redox balance says that the total redox influx to the atmosphere (i.e., surface emission) should be balanced by the total redox outflux from the atmosphere (i.e., atmospheric escape, dry deposition, and wet deposition), otherwise the atmosphere is being oxidized or reduced. The redox balance is equivalent to the conservation of total budget of hydrogen in the atmosphere, and is equivalent to the conservation of the total number of electrons in the atmosphere. One might consider the redox balance to be redundant. The redox balance is redundant if the lower and upper boundary conditions for all species are specified in fluxes (and not in mixing ratios). In many applications, however, it is useful to use the Type-1 lower boundary condition, which specifies mixing ratio. In that case, the mass conservation would not necessarily guarantee the redox balance. After all, an imbalance in the redox budget for a steady-state solution indicates either bugs in chemistry kinetics or unphysical boundary conditions. We explicitly check the redox balance for all our model outputs.

\section{Model Validation}

We validate the photochemistry model by computing atmospheric compositions of current Earth and Mars. Due to the one-dimensional nature of the photochemistry model, we intend to compare with global averages of observations, if available. In most cases, however, the global average is only an order-of-magnitude representation of chemical composition of Earth's and Mars' atmospheres, because of strong diurnal and spatial variations. We should therefore expect our one-dimensional chemical-transport model to match quantitatively with the global average values to within a factor of two or three. For Mars, vertical profiles of most trace gas content have not yet been measured. We therefore compare our results with other photochemistry model outputs from the literature.

\subsection{Transport-Only Model for an Earth-like Atmosphere}

As a test of transport-related schemes of the photochemical model, we first consider a transport-only case. In theory, for species whose removal timescales are significantly larger than their transport timescales, such as \ce{CO2} in Earth's atmosphere, their mixing ratios do not change with altitude, i.e., they are well-mixed.

\begin{figure}[h]
\begin{center}
 \includegraphics[width=0.45\textwidth]{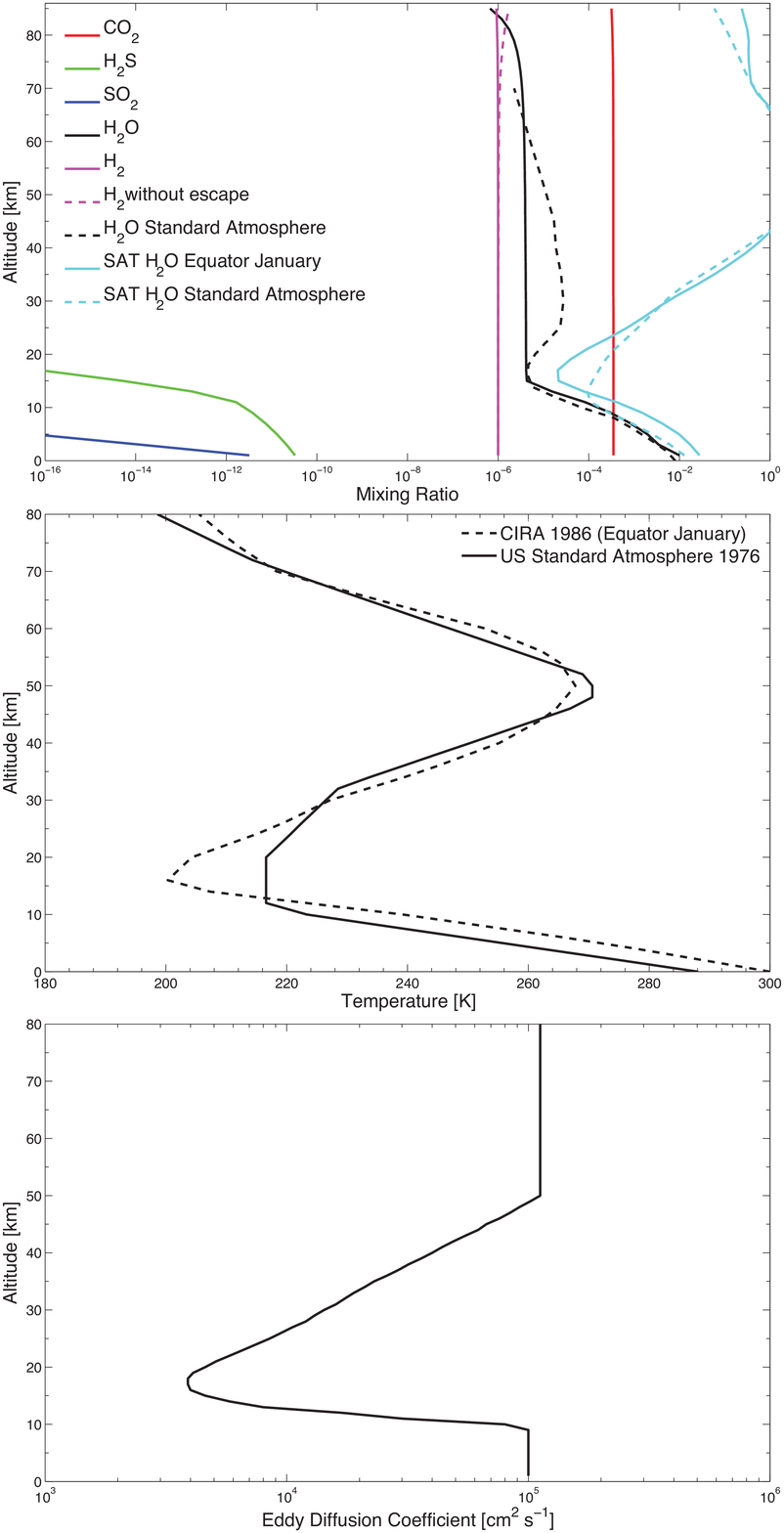}
 \caption{Validation of the transport and condensation schemes of our photochemistry code.
 Only vertical transport via eddy diffusion, molecular diffusion (for \ce{H2} only), dry deposition, wet deposition, condensation of water vapor, diffusion-limited escape of \ce{H2}, and photolysis of \ce{CO2}, \ce{H2O}, \ce{H2}, \ce{SO2} and \ce{H2S} are considered for an Earth-like atmosphere. We omit all chemical kinetics to exclusively test the transport-related schemes. \ce{H2S} and \ce{SO2} are the only sulfur compounds in this test and they are considered as removed once photo-dissociated or deposited.
 The atmosphere has the temperature-pressure profile of COSPAR International Reference Atmosphere (CIRA) 1986 at the equator in January, as shown by the dashed line on the middle panel, and major constituents of 20\% \ce{O2} and 80\% \ce{N2}. Also on the middle panel we plot the temperature profile of US Standard Atmosphere 1976 for comparison.
 We adopt the eddy diffusion coefficient empirically derived from vertical profiles of trace gases (Massie \& Hunten, 1981), shown in the lower panel.
 The boundary conditions are set as follows: \ce{CO2} mixing ratio at the surface 350 ppm; \ce{H2O} mixing ratio at the surface 0.01; \ce{H2} mixing ratio at the surface 1 ppm; \ce{SO2} surface emission flux $9.0\times10^9$ cm$^{-2}$ s$^{-1}$; \ce{H2S} surface emission flux $2.0\times10^8$ cm$^{-2}$ s$^{-1}$.
 Water vapor in the atmosphere is limited by condensation.
 Steady-state mixing ratios are shown on the left panel, and the solid and dashed lines for \ce{H2} show the situations with and without escape.
 Light blue lines on the left panel are saturation mixing ratios of water vapor with two temperature profiles. We verify that long-lived species, such as \ce{CO2}, are well-mixed in the atmosphere.
 }
 \label{Transport}
  \end{center}
\end{figure}

We turn off the chemical network deliberately and compute a model using Earth's temperature profile to test the eddy diffusion transport, molecular diffusion transport, condensation, and rainout schemes of our chemical-transport model. A valid transport model should preserve hydrostatic equilibrium, predict well mixed mixing ratios for long-lived species, and predict a mixing ratio gradient for species that is rapidly removed in the atmosphere. We have verified that the transport scheme, although written in terms of number density rather than mixing ratio, preserves hydrostatic equilibrium. The fact that the code maintains hydrostatic equilibrium indicates that the transport scheme is numerically correct. We verify that long-lived species are well mixed, such as \ce{CO2} throughout the atmosphere, as shown in Figure \ref{Transport}. We assign a mixing ratio for \ce{CO2} at the surface and the code is able to predict a well-mixed vertical profile. We also verify that for \ce{H2} molecular diffusion tends to increase the mixing ratio when the molecular diffusion coefficient is comparable with the eddy diffusion coefficient (i.e., the homopause; see Figure \ref{Transport}). With a diffusion-limited escape, the effect of molecular diffusion and the effect of escape on the \ce{H2} vertical profile near the homopause largely cancel out (Figure \ref{Transport}), consistent with the definition of diffusion-limited escape. We also show the behaviour of \ce{H2S}, a poorly mixed species, to compare with well-mixed CO2. In this simplified model the only way to remove \ce{H2S} is photolysis, which requires photons with wavelengths shorter than 260 nm. Most of the photons in this wavelength range are effectively shielded by the major species \ce{O2}, so that removal rate of \ce{H2S} by photolysis near the surface is smaller than that in the stratosphere (see Figure \ref{Transport}).

Water vapor content in a rocky exoplanet's atmosphere is controlled by the water reservoir at the planetary surface, vertical transport, and condensation. In our model, the mixing ratio of water vapor at the surface is assigned according to appropriate temperature-dependent relative humidity. Water vapor can be transported up into the atmosphere by eddy diffusion, and as the temperature decreases the atmosphere may become supersaturated in water as the altitude increases. For an Earth-like planet atmosphere, it is appropriate to assume a water vapor mixing ratio of 0.01 at the surface (which corresponds to a relative humidity of about 60\%). The condensation scheme becomes effective when water vapor is supersaturated, which keeps the water vapor profile along the saturation profile. Such an approach to computing the atmosphere water vapor content is commonly adopted by other previously described photochemistry models of terrestrial planets (e.g., Nair et al. 1994; Yung \& Demore 1999; Zahnle et al. 2008).

The one-dimensional transport model tends to saturate the tropopause and over-predict the amount of water vapor in the stratosphere. In fact, when using the US Standard Atmosphere 1976 as the temperature profile, the tropopause temperature is 217 K, much warmer than the required ``cold trap" temperature ($\sim200$ K) in order to maintain a dry stratosphere (water vapor mixing ratio in a few ppm). Common 1-D transport-condensation schemes may over-predict the amount of water vapor above the cold trap by a few orders of magnitude, and therefore over-predict the strength of \ce{HO_x} cycle in the stratosphere. This is a well known problem in modeling Earth's atmosphere, mainly due to spatial and temporal variability of tropopause temperatures (K. Emmanuel, 2012, private communication). Tropospheric convection is most effective in transporting water vapor in tropics, where the tropopause temperature is the lowest and the number density of water vapor at the tropopause is the lowest compared with other latitudes. It is therefore likely that the tropopause is highly unsaturated as a global average. We use the temperature profile of the equatorial region in January (see Figure \ref{Transport}), and limit the water vapor saturation ratio to within 20\%. In this way we reproduce the water vapor profile of the US Standard Atmosphere 1976, which has a dry stratosphere (Figure \ref{Transport}). The classic Manabe-Wetherald relative humidity profile for 1-D photochemistry models of Earth's atmosphere also has 20\% relative humidity at the tropopause (Manabe \& Wetherald 1967).

As a result of water vapor condensation and rain, soluble gases are removed from the atmosphere by wet deposition (rainout). We expect to see a decreasing slope in mixing ratio of the species being rained out, which depends on the rainout rate, itself dependent on the solubility of the species. For example, \ce{SO2} is much more soluble than \ce{H2S}, so that \ce{SO2} is more effectively rained out in the troposphere and cannot accumulate in the troposphere (see Figure \ref{Transport}).

\subsection{Present-Day Earth}

We compute a photochemical model to simulate the present-day Earth, and compare our results with globally averaged measurements. The most important photochemical process in Earth's atmosphere is formation of the ozone layer, which leads to the temperature inversion in the stratosphere (e.g., Seinfeld \& Pandis 2006). On the other hand, the tropospheric chemistry is mostly controlled by the hydrological cycle and surface emission of \ce{CH4}, \ce{NO_x} (i.e., \ce{NO} and \ce{NO2}) and \ce{SO2}, and involves coupled processes of photochemistry and deposition. The reproduction of current Earth atmospheric composition is therefore a comprehensive validation of various aspects of the photochemistry model.

We start with a nominal 80\% \ce{N2} and 20\% \ce{O2} composition, having well-mixed \ce{CO2} of 350 ppm. We assume the water vapor mixing ratio at the surface is 0.01 (i.e., type-1 lower boundary condition), and let the water vapor be transported in the atmosphere and condense near the tropopause. The temperature profiles are assumed to be the 1976 US Standard Atmosphere (subsequently referred to as ``Model A") or a reference tropical atmosphere in January of 1986 from COSPAR\footnote{Committee on Space Research} International Reference Atmosphere (CIRA, subsequently referred to as ``Model B"; ), and the eddy diffusion coefficient is adopted from Massie \& Hunten (1981), who have derived it from vertical profiles of trace gases. We model the atmosphere from 0 to 86 km altitude as 43 equally spaced layers. The incoming solar radiation is cut off for wavelengths shorter than 100 nm to account for the thermosphere absorption. The photolysis rates at the top layer of the atmosphere are tabulated in Table \ref{Photolysis}. Key trace species are assigned emission rates from the surface according to typical global values (Seinfeld \& Pandis 2006). Emission rates and dry deposition velocities of these species are tabulated in Table \ref{EarthEmission}.

\begin{table}[h]
\caption{ Photochemistry model validation. For the surface emission and dry deposition velocities, measured surface mixing ratios are compared to our models using the 1976 US Standard Atmosphere (Model A) and using the reference tropical temperature profile (Model B).}
\begin{center}
\begin{tabular}{l|ll|lll}
  \hline
  Species & Surface Emission\tablenotemark{a} & $V_{\rm DEP}$ & \multicolumn{2}{l}{Surface Mixing Ratio} & \\
  & (molecule cm$^{-2}$ s$^{-1}$) & (cm s$^{-1}$) & Measured\tablenotemark{e} & Model A & Model B\\
  \hline
  \ce{CO}       & $3.7\times10^{11}$ & 0.03\tablenotemark{b} & 40-200 ppb & 113 ppb & 101 ppb \\
  \ce{CH4}      & $1.4\times10^{11}$ & 0.03\tablenotemark{b} & 700-1745 ppb\tablenotemark{f} & 1939 ppb & 1235 ppb \\
  \ce{NH3}      & $7.7\times10^{9}$  & 1\tablenotemark{d} & 0.1-10 ppb  & 0.24 ppb & 0.24 ppb \\
  \ce{N2O}      & $1.0\times10^{9}$  & 0\tablenotemark{b} & 276-315 ppb\tablenotemark{f}  & 302 ppb & 290 ppb \\
  \ce{NO}       & $7.0\times10^{9}$  & 0.016\tablenotemark{b} & 0.02-10 ppb\tablenotemark{g} & 0.024 ppb & 0.025 ppb\\
  \ce{SO2}      & $9.0\times10^{9}$  & 1\tablenotemark{c} & 30-260 ppt & 237 ppt & 239 ppt \\
  \ce{OCS}      & $5.0\times10^{7}$  & 0.01\tablenotemark{d} & 510 ppt & 188 ppt & 185 ppt\\
  \ce{H2S}      & $2.0\times10^{8}$  & 0.015\tablenotemark{c} & 1-13 ppt & 3.92 ppt & 3.62 ppt \\
  \ce{H2SO4}    & $7.0\times10^{8}$ & 1\tablenotemark{d} & 5-70 ppt\tablenotemark{h} & 127 ppt\tablenotemark{h} & 126.7 ppt\tablenotemark{h} \\
  \hline
\end{tabular}
\tablenotetext{a}{Typical globally averaged emission rates are taken from Seinfeld \& Pandis (2006).}
\tablenotetext{b}{Typical dry deposition velocities are taken from the compilation of Hauglustaine et al. (1994).}
\tablenotetext{c}{Typical dry deposition velocities are taken from the compilation of Sehmel (1980).}
\tablenotetext{d}{Dry deposition velocities are assumed by considering the solubility and reactivity of gases (see Appendix \ref{A_DD}).}
\tablenotetext{e}{Mixing ratios at the surface are taken from Seinfeld \& Pandis (2006).}
\tablenotetext{f}{Ranges are from the preindustrial mixing ratio to the present-day value.}
\tablenotetext{g}{Ranges are mixing ratio of \ce{NO_x} at the planetary boundary layer.}
\tablenotetext{h}{Mixing ratio of \ce{SO4^2-} in both gaseous and aqueous phases.}
\end{center}
\label{EarthEmission}
\end{table}

\begin{figure}[h]
\begin{center}
 \includegraphics[width=1.0\textwidth]{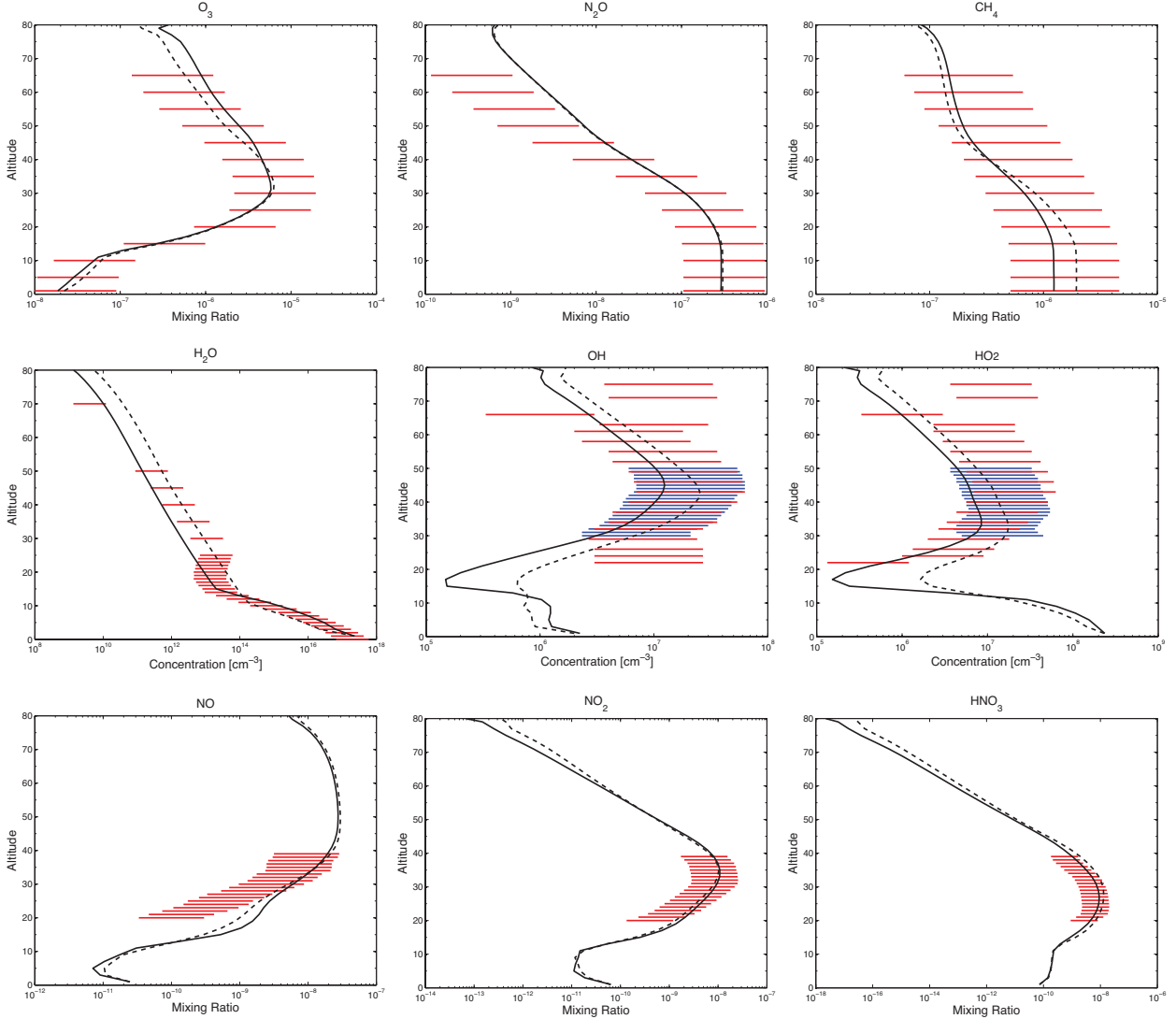}
 \caption{Profiles of key molecules in Earth's atmosphere predicted by our photochemistry models using the US Standard Atmosphere 1976 for mid-latitudes (Model A; dashed lines)and the reference tropical temperature profile (Model B; solid lines), compared with selected observations (horizontal lines). We adopt the eddy diffusion coefficient empirically derived from vertical profiles of trace gases (Massie \& Hunten, 1981), as shown in Figure \ref{Transport}. The error bars of observations are uniformly set to 1 order of magnitude to account for diurnal and spatial variations.
 The terrestrial data are: (1) globally averaged mixing ratios of \ce{O3}, \ce{N2O} and \ce{CH4} compiled by Massie \& Hunten (1981); (2) number density of \ce{H2O} from the US Standard Atmosphere 1976 for mid-latitudes; (3) number densities of \ce{OH} and \ce{HO2} from balloon observations in Fairbanks, AK in 1997 (blue lines; Jucks et al. 1998); (4) number densities of \ce{OH} and \ce{HO2} from AURA satellite observations using the Microwave Limb Sounder that are zonally averaged over a latitude interval of 20$^{\circ}$ for a period of 15 days (Pickett et al. 2006); (5) mixing ratios of \ce{NO}, \ce{NO2} and \ce{HNO3} from balloon observations at Lat$\sim$35$^{\circ}$N in 1993 (Sen et al. 1998).
 }
 \label{EarthCompare}
  \end{center}
\end{figure}

Our models correctly simulate the photochemistry in Earth's stratosphere. First, the formation of the ozone layer is correctly predicted. The modeled vertical profiles and amounts of ozone are consistent with the globally averaged measurements, as shown in Figure \ref{EarthCompare}. Second, the nitrogen oxide cycle and the \ce{HO_x} cycle are properly simulated. In Earth's stratosphere, chemical species are rapidly converted to each other within the \ce{NO_x} group and the \ce{HO_x} group, and both cycles lead to catalytic destruction of ozone (e.g., Seinfeld \& Pandis 2006). We reproduce the vertical profiles of \ce{NO_x} and \ce{HO_x} in the stratosphere (see Figure \ref{EarthCompare}), qualitatively consistent with mid-latitude measurements. It appears that the models tend to over-predict the amounts of \ce{NO} in the lower stratosphere. Third, for relatively long lived species in the stratospheres like \ce{N2O} and \ce{CH4}, our models produce correct altitude gradient of mixing ratios.

The key element of the tropospheric chemistry is the production of the hydroxyl radical \ce{OH}, because \ce{OH} is the major removal pathway for most species emitted from the surface. It has been established that \ce{OH} in Earth's troposphere is produced by reactions between \ce{O(^1D)} and \ce{H2O}, and \ce{O(^1D)} is produced by photolysis of \ce{O3} (Seinfeld \& Pandis, 2006). In turn, the main source of tropospheric ozone is the \ce{NO_x} cycle, with non-neglible effects of \ce{HO_x} cycle and hydrocarbon chemistry as well. We find the surface ozone mixing ratio is 19 ppb and 22 ppb for Model 1976 and Model 1986 respectively, and the surface \ce{OH} number density is $2.0\times10^{6}$ cm$^{-3}$ for Model 1976 and $2.2\times10^{6}$ cm$^{-3}$ for Model 1986, very close to the commonly adopted \ce{OH} number density of $1.0\times10^{6}$ cm$^{-3}$. We also confirm that the steady-state mixing ratios of major trace gases near the surface are consistent with ground measurements (see Table \ref{EarthEmission}). The main removal mechanism of \ce{CO}, \ce{CH4}, and \ce{NH3} is through reactions of \ce{OH}, and \ce{N2O} is long-lived in the troposphere and is transported up to the stratosphere. Considering the significant spatial and temporal variability of the amount of \ce{OH}, we conclude that our photochemistry model correctly computes the chemistry in Earth's troposphere.

Our photochemistry model correctly treats the sulfur chemistry in Earth's atmosphere (see Table \ref{EarthEmission}). All sulfur-bearing emission, if not deposited, is oxidized in multiple steps in the troposphere and eventually converted into sulfate. Sulfite (\ce{S^{4+}}) and sulfate (\ce{S^{6+}}) are soluble in water and effectively removed from the atmosphere by rainout. Our photochemistry models simulate these processes, and find steady-state mixing ratios of sulfur-bearing species (e.g., \ce{H2S} and \ce{SO2}) consistently with the ground measurements. The models also predict the saturation of sulfate and the formation of sulfate aerosols as expected. The modeled mixing ratio of sulfate is slightly larger than observations, which might be related to the fact that we do not consider the hydration of \ce{H2SO4}, nor the sulfate-facilitated cloud formation in our models.

In summary, we validate our photochemistry model by reproducing the chemical composition of current Earth's atmosphere (see Table \ref{EarthEmission} and Figure \ref{EarthCompare}). The photochemistry model successful in predicting the formation of the ozone layer, treating key chemical cycles in both the stratosphere and the troposphere, computing oxidation of hydrocarbon, ammonia and sulfur-bearing species in the troposphere, and transporting long-lived species from the troposphere to the stratosphere. These aspects involve all chemical kinetics, photolysis, and transport processes, which not only verify that our photochemistry model is suitable for applications in oxidizing atmospheres, but also allows our model to be applied to other atmospheric scenarios.

\subsection{Present-Day Mars}

\begin{figure}[h]
\begin{center}
 \includegraphics[width=0.9\textwidth]{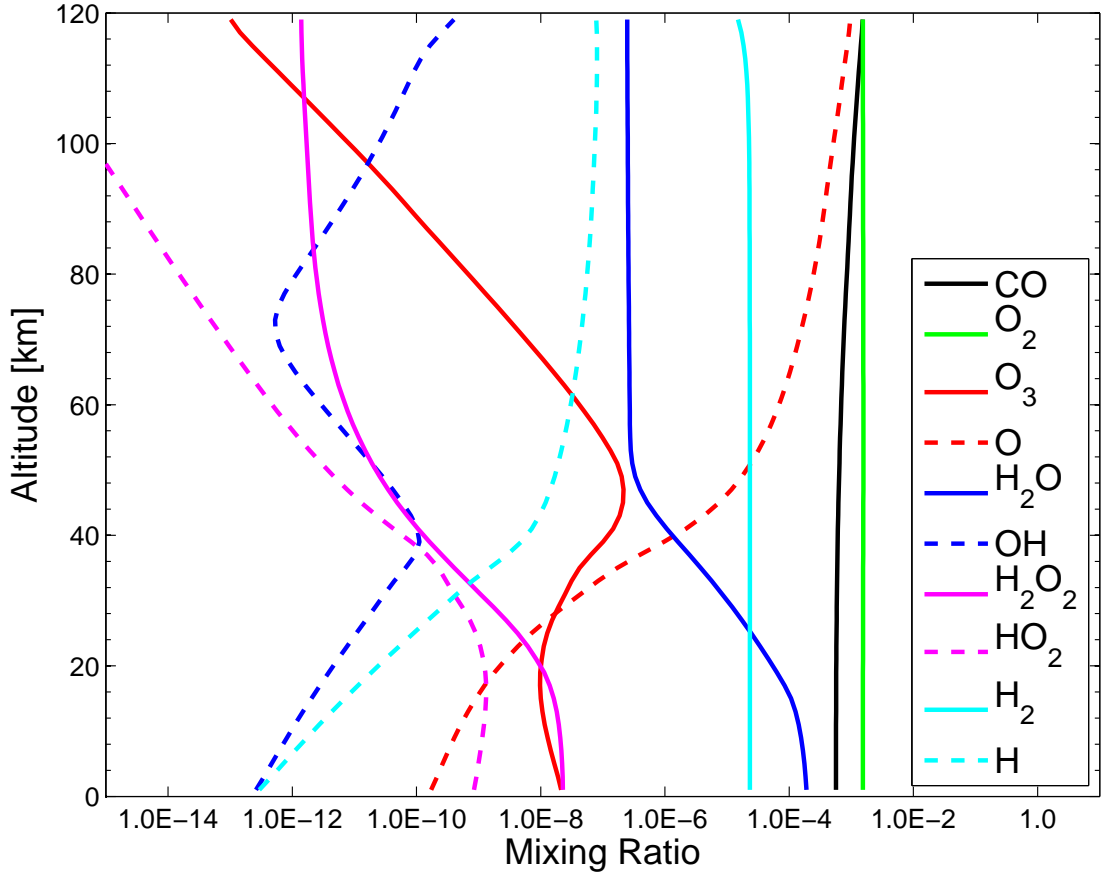}
 \caption{Profiles of key molecules in Mars' atmosphere predicted by our photochemistry model.
 The chemical-transport model is computed with a 2-km grid from 0 to 120 km.
 For the UV flux computation, unattenuated solar flux at 1.524 AU and surface albedo of 0.1 are used.
 UV cross sections are computed at 200 K when their temperature dependencies are available from laboratory measurements.
 All C, H, O, N species and relevant reactions are computed. Ter-molecular reaction rates are multiplied by 2.5 to account for \ce{CO2} being the third body (e.g., Zahnle et al. 2008).
 The eddy diffusion coefficient is assumed to have a profile as equation (\ref{eddy}) with $K_T = 10^6$ cm$^2$ s$^{-1}$, $K_H = 10^8$ cm$^2$ s$^{-1}$ and $z_T = 20$ km.
 The atmospheric temperature profile is assumed as Zahnle et al. (2008): surface temperature 210 K, lapse rate of 1.4 K km$^{-1}$ untill 50 km and then isothermal of 140 km up to 120 km.
 The relative humidity at the surface is fixed at 0.19, and the water vapor is transported up into atmosphere by eddy mixing, limited by condensation. The maximum saturation of water vapor of 60\% is imposed, which gives a column number density of 9.8 precipitable microns. Wet deposition is reduced compared to Earth, with the rainout factor assumed to be 0.01 in the model.
 }
 \label{MarsModel}
  \end{center}
\end{figure}

We validate our photochemistry model by simulating the atmosphere of present-day Mars. The current atmosphere of Mars is a thin \ce{CO2}-based atmosphere and its bulk chemical composition is known from extensive ground-based observations (see Krasnopolsky 2006 and references therein). We validate our photochemistry code by reproducing the observed mixing ratios of \ce{CO} and \ce{O3}, and comparing with previous results regarding Martian atmosphere photochemistry. Key parameters and results of the present-Mars atmosphere model are shown in Table \ref{MarsEmission} and Figure \ref{MarsModel}. We emphasize that it is important to use temperature-dependent UV cross sections for modeling the photolysis and UV penetration in the Martian environment (see Anbar et al. 1993 for an error analysis of temperature-dependence of \ce{CO2} in the Martian atmosphere).

The most important aspect of Martian atmospheric photochemistry is the stabilization of the \ce{CO2} atmosphere by \ce{H}, \ce{OH}, and \ce{HO2} (commonly referred to as odd hydrogens; Nair et al. 1994; Krasnopolsky 2006; Zahnle et al. 2008). Odd hydrogens, as catalysts, facilitate the recombination of photochemically produced \ce{CO} and \ce{O2} and maintain the \ce{CO2} dominance of the Mars' atmosphere (e.g. Nair et al. 1994; Krasnopolsky 2006; Zahnle et al. 2008). A pure \ce{CO2} atmosphere is unstable against photolysis, because \ce{CO2} can only be restored with a slow three-body reaction \ce{CO + O + M -> CO2 + M } (e.g., Yung \& DeMore 1999). Odd hydrogen species, including \ce{H}, \ce{OH} and \ce{HO2}, are produced by photolysis of water vapor; and trace amounts of odd hydrogens can effectively stabilize the \ce{CO2} dominant atmosphere (e.g. Nair et al. 1994; Krasnopolsky 2006; Zahnle et al. 2008). Nonetheless, one-dimensional photochemistry models tend to over-address the problem, predicting amounts of \ce{CO}, \ce{O2} and \ce{O3} several times smaller than the observed values (e.g. Nair et al. 1994; Krasnopolsky 2006). By considering a slow dry deposition of \ce{H2O2} and \ce{O3} to the surface, Zahnle et al. (2008) are able to predict the amount of \ce{CO}, \ce{O2} and \ce{H2} that match observations. We follow the assumptions of Zahnle et al. (2008), and confirm the finding of appropriate amounts of \ce{O2}, \ce{CO} and \ce{H2} (see Table \ref{MarsEmission} and Figure \ref{MarsModel}). We therefore reproduce the photochemical stability of the current Martian atmosphere.

The timescale of \ce{CH4} removal can also be used for model validation. Rapid variation (over several years) of the amount of \ce{CH4} in Mars' atmosphere has been reported (e.g., Mumma et al. 2009), but the modeling of coupled general circulation and gas-phase chemistry find no known gas-phase chemistry path that allow such a rapid removal (e.g., Lef\`evre \& Forget 2009). Based on our fiducial model of Mars' atmosphere, the loss timescale for \ce{CH4} is computed to be about 240 years, within the same order of the magnitude of Lef\`evre \& Forget (2009).

\begin{table}[h]
\caption{Photochemistry model validation. For the upper boundary flux and dry deposition velocities, measured mixing ratios of major trace gases on Mars are compared to our model.
}
\begin{center}
\begin{tabular}{l|ll|lll}
  \hline
  Species & Upper Boundary Flux\tablenotemark{a} & $V_{\rm DEP}$\tablenotemark{b} & \multicolumn{2}{l}{Column Averaged Mixing Ratio} & \\
  & (molecule cm$^{-2}$ s$^{-1}$) & (cm s$^{-1}$) & Measured\tablenotemark{e} & Modeled \\
  \hline
  \ce{O2}       & 0 & 0 & 1200-2000 ppm & 1545 ppb \\
  \ce{CO}      & -$2.0\times10^{7}$ & 0 & 800 ppm & 572 ppb \\
  \ce{H2}        & $3.6\times10^{8}$ & 0 & 17 ppm & 23 ppb \\
  \ce{H2O2}       & 0 & 0.02 & $0\sim40$ ppb\tablenotemark{f} & 18 ppb \\
  \ce{O3}		& 0 & 0.02 & $0\sim120$ ppb\tablenotemark{f} & 18 ppb \\
    \hline
\end{tabular}
\tablenotetext{a}{ \ce{H2} upper boundary flux is computed from diffusion-limited escape velocity after finding the steady state.
Besides tabulated values, influx of \ce{N} of $2.0\times10^{6}$ molecule cm$^{-2}$ s$^{-1}$, \ce{NO} of $2.0\times10^{7}$ molecule cm$^{-2}$ s$^{-1}$, and outflux of \ce{O} of $2.0\times10^{7}$ molecule cm$^{-2}$ s$^{-1}$, are considered as input to the model.
}
\tablenotetext{b}{
Dry deposition velocity is assumed to be 0.02 for reactive species, including \ce{H}, \ce{O}, \ce{O(^1D)}, \ce{O3}, \ce{OH}, \ce{HO2}, \ce{H2O2}, \ce{CHO2}, \ce{CH2O2}, \ce{CH3O2}, \ce{CH4O2}, \ce{NH3}, \ce{NO3}, \ce{N2O5}, \ce{HNO3}, and \ce{HNO4} (e.g., Zahele et al. 2008). Dry deposition velocity is assumed to be zero for all other species.
}
\tablenotetext{e}{Mars data are from the compilation of Krasnopolsky (2006).}
\tablenotetext{f}{ Mixing ratios of \ce{O3} and \ce{H2O2} have significant diurnal, seasonal and latitudinal variations.}
\end{center}
\label{MarsEmission}
\end{table}

In summary, we validate our photochemistry model by reproducing the atmospheric compositions of current Earth and Mars. We find that the model gives consistent results compared to observations and previous photochemistry models. All physical and chemical processes including photolysis, chemical reactions, transports, condensation, and deposition are rigorously tested in these examples.

\section{Exoplanet Benchmark Cases}

We now present three benchmark atmospheric scenarios of rocky exoplanets and summarize the effects of key photochemistry processes. The goal is to provide baseline models to assess the stability of molecules in different kinds of atmospheres in order to: identify the dominant stable molecules; calculate the lifetime of spectrally significant gases; and identify the amounts of the main reactive species that control molecule lifetimes. The benchmark cases are also intended to serve as the test cases for independent photochemistry models for rocky exoplanet atmospheres.

The key to assess molecular stability is the oxidation state of the atmosphere, because the main reactive species are linked to the oxidation state of the atmosphere. In an oxidizing atmosphere, \ce{OH} and \ce{O} are created by photochemistry and are the main reactive radicals. In a reducing atmosphere, \ce{H}, also created by photochemistry, is the main reactive species. Although we expect the atmospheric composition will be highly varied, based on the nearly continuous range of masses and orbits of exoplanets, we believe that the primary dimension of chemical characterization for terrestrial exoplanet atmospheres is their oxidation states. For the benchmark cases, therefore, we have chosen three endmembers in terms of atmospheric oxidizing power. The scenarios are a reducing (90\%\ce{H2}-10\%\ce{N2}) atmosphere, a weakly oxidizing \ce{N2} atmosphere ($>99$\%\ce{N2}), and a highly oxidizing (90\%\ce{CO2}-10\%\ce{N2}) atmosphere.  We consider Earth-like volcanic gas composition that consists of \ce{CO2}, \ce{H2}, \ce{SO2}, \ce{CH4}, and \ce{H2S}, with emission rates comparable to current Earth. We assume that the planet surface has a substantial fraction of its surface covered by a liquid water ocean so that water is transported from the surface and buffered by the balance of evaporation/condensation. Key assumptions of the parameters of the three atmospheric scenarios are summarized in Table \ref{AtmosPara} and rationals of important model parameters are given in \S~\ref{ParaPhysics}.

\begin{table}[htdp]
\tiny
\caption{Basic parameters for the terrestrial exoplanet benchmark scenarios. The benchmark scenarios are \ce{H2}, \ce{N2}, \ce{CO2} atmospheres on habitable terrestrial exoplanets with Earth-like volcanic emissions. Note that we do not consider any known biosignature gas emission or biotic contribution to the dry deposition velocities. }
\begin{center}
\begin{tabular}{llll}
\hline\hline
Parameters & Reducing & Weakly Oxidizing & Highly Oxidizing\\
\hline
Main component & 90\%\ce{H2}, 10\%\ce{N2}  & $>99$\%\ce{N2}  & 90\%\ce{CO2}, 10\%\ce{N2}  \\
Mean molecular mass & 4.6  & 28  & 42.4  \\
\hline
\multicolumn{4}{l}{\it Planetary parameters}\\
Stellar type & G2V & G2V & G2V \\
Semi-major axis & 1.6 AU & 1.0 AU & 1.3 AU \\
Mass & $M_{\earth}$ & $M_{\earth}$ &  $M_{\earth}$ \\
Radis & $R_{\earth}$ & $R_{\earth}$ & $R_{\earth}$ \\
\hline
\multicolumn{4}{l}{\it Temperature profile}\\
Surface temperature & 288 K & 288 K & 288 K \\
Surface pressure & $10^5$ Pa & $10^5$ Pa & $10^5$ Pa \\
Tropopause altitude & 120 km & 13.4 km & 8.7 km \\
Temperature above tropopause & 160 K & 200 K & 175 K \\
Maximum altitude &  440 km & 86 km & 51 km \\
\hline
\multicolumn{4}{l}{\it Eddy diffusion coefficient}\\
In the convective layer & $6.3\times10^5$ cm$^2$ s$^{-1}$ & $1.0\times10^5$ cm$^2$ s$^{-1}$ & $6.8\times10^4$ cm$^2$ s$^{-1}$ \\
Minimum & $2.5\times10^4$ cm$^2$ s$^{-1}$ & $3.9\times10^3$ cm$^2$ s$^{-1}$ &$2.7\times10^3$ cm$^2$ s$^{-1}$ \\
Altitude for the minimum & 107 km & 17.0 km & 11.6 km \\
Near the top of atmosphere & $7.1\times10^5$ cm$^2$ s$^{-1}$ & $1.1\times10^5$ cm$^2$ s$^{-1}$ & $7.6\times10^4$ cm$^2$ s$^{-1}$ \\
\hline
\multicolumn{4}{l}{\it Water and rainout}\\
Liquid water ocean & Yes & Yes & Yes \\
Water vapor boundary condition & $f(\ce{H2O})=0.01$ & $f(\ce{H2O})=0.01$ & $f(\ce{H2O})=0.01$ \\
Rainout rate\tablenotemark{a} & Earth-like & Earth-like & Earth-like \\
\hline
\multicolumn{4}{l}{\it Gas emission\tablenotemark{b} }\\
\ce{CO2} & $3\times10^{11}$ cm$^{-2}$ & $3\times10^{11}$ cm$^{-2}$ s$^{-1}$ & N/A\\
\ce{H2} & N/A & $3\times10^{10}$ cm$^{-2}$ s$^{-1}$ & $3\times10^{10}$ cm$^{-2}$ s$^{-1}$ \\
\ce{SO2} & $3\times10^9$ cm$^{-2}$ & $3\times10^9$ cm$^{-2}$  & $3\times10^9$ cm$^{-2}$  \\
\ce{CH4} & $3\times10^8$ cm$^{-2}$ s$^{-1}$ & $3\times10^8$ cm$^{-2}$  & $3\times10^8$ cm$^{-2}$ \\
\ce{H2S} & $3\times10^8$ cm$^{-2}$ & $3\times10^8$ cm$^{-2}$  & $3\times10^8$ cm$^{-2}$  \\
\hline
\multicolumn{4}{l}{\it Dry deposition velocity\tablenotemark{c} }\\
\ce{H2} &\multicolumn{3}{l}{0} \\
\ce{CH4} &\multicolumn{3}{l}{0}\\
\ce{C2H6} &\multicolumn{3}{l}{ $1.0\times10^{-5}$ (Assumed) }\\
\ce{CO} &\multicolumn{3}{l}{$1.0\times10^{-8}$ cm s$^{-1}$ (Kharecha et al. 2005) } \\
\ce{CH2O} &\multicolumn{3}{l}{0.1 cm s$^{-1}$ (Wagner et al. 2002) } \\
\ce{CO2} &\multicolumn{3}{l}{$1.0\times10^{-4}$ cm s$^{-1}$ (Archer 2010) }\\
\ce{O2} &\multicolumn{3}{l}{0}\\
\ce{O3} & \multicolumn{3}{l}{0.4 cm s$^{-1}$ (Hauglustaine et al. 1994) } \\
\ce{H2O2} & \multicolumn{3}{l}{0.5 cm s$^{-1}$ (Hauglustaine et al. 1994) } \\
\ce{H2S} & \multicolumn{3}{l}{0.015 cm s$^{-1}$ (Sehmel 1980) }\\
\ce{SO2} & \multicolumn{3}{l}{1.0 cm s$^{-1}$ (Sehmel 1980) } \\
\ce{S8(A)} & \multicolumn{3}{l}{0.2 cm s$^{-1}$ (Sehmel 1980) } \\
\ce{H2SO4(A)} & \multicolumn{3}{l}{0.2 cm s$^{-1}$ (Sehmel 1980) }\\
\hline\hline
\end{tabular}
\tablenotetext{a}{Rainout rates for \ce{H2}, \ce{CO}, \ce{CH4}, \ce{C2H6}, and \ce{O2} are generally assumed to be zero to simulate an ocean surface saturated with these gases on an abiotic exoplanet. }
\tablenotetext{b}{The volcanic gas emission rates from the planetary surface are assigned for each model scenario. \ce{H2O} emission is not explicitly considered because the surface has a large water reservoir, i.e., an ocean.}
\tablenotetext{c}{We here list the dry deposition velocities (with references) for emitted gases and their major photochemical byproducts, and dry deposition velocities that are important for the mass and redox balance of the atmosphere. Dry deposition velocities are assumed to be identical for the three scenarios. \ce{C2H6} dry deposition velocity is assumed to take into account the loss of carbon due to organic haze formation and deposition. \ce{CO2} dry deposition velocity is estimated from a 10,000-year lifetime that matches with the lifetime of silicate weathering on Earth (Archer 2010; S. Solomon, private commonication). }
\end{center}
 \label{AtmosPara}
\end{table}

\begin{figure}[h]
\begin{center}
 \includegraphics[width=0.4\textwidth]{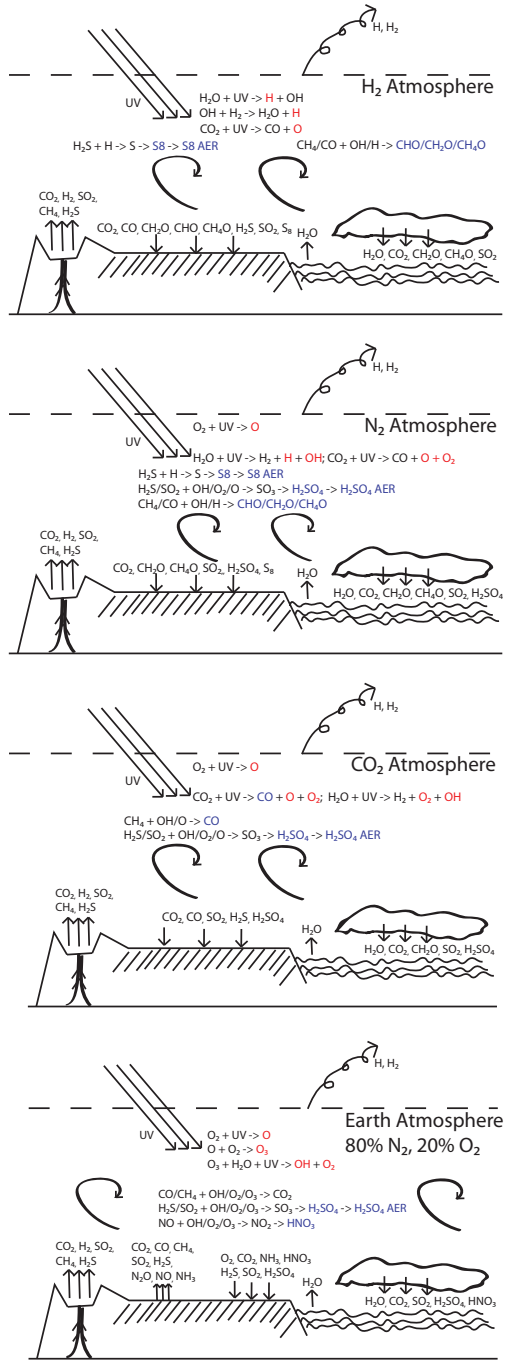}
 \caption{ Schematic illustrations of key non-equilibrium processes in the 3 scenarios of rocky exoplanet atmospheres considered in this paper (Table \ref{AtmosScenario}), in comparison with the current Earth. From top to bottom, the four panels correspond to the \ce{H2}, \ce{N2}, \ce{CO2} atmospheres, and the atmosphere of Earth. The red color highlights the reactive radicals in each atmospheric scenario, and the blue color highlights the major photochemical products in the atmosphere.
 }
 \label{Schematic}
  \end{center}
\end{figure}

We now describe the key results of the three atmosphere scenarios, for the 1-bar atmosphere on an Earth-size and Earth-mass habitable rocky exoplanet around a Sun-like star.  We also describe qualitatively the behaviors of the atmospheres in the habitable zone of a quiet M dwarf, based on our numerical explorations. Mixing ratios of emitted gases, photochemical products, and reactive species in the three scenarios are tabulated in Table \ref{AtmosScenario} and shown in Figure \ref{Benchmark}. Schematic illustrations of key non-equilibrium processes in the three scenarios are shown in Figure \ref{Schematic}. We start with a qualitative overview of the key results, and then present the chemical properties of the three benchmark scenarios.

\subsection{General Results}

We here list our main findings on general chemistry properties of atmospheres on habitable terrestrial exoplanets.

Atomic hydrogen (\ce{H}) is a more abundant reactive radical than hydroxyl radical (\ce{OH}) in anoxic atmospheres. Atomic hydrogen is mainly produced by water vapor photodissociation; and in anoxic atmospheres the main ways to remove atomic hydrogen is its recombination and reaction with \ce{CO}. This is in contrast to oxygen-rich atmospheres (e.g., current Earth's atmosphere) in which \ce{H} is quickly consumed by \ce{O2}. As a result, removal of a gas by \ce{H} is likely to be an important removal path for trace gases in an anoxic atmosphere. Atomic oxygen is the most abundant reactive radical in \ce{CO2} dominated atmospheres. Due to the photochemical origin of the reactive species including \ce{H}, \ce{OH}, and {O}, their abundances in the atmosphere around a quiet M dwarf are 2 orders of magnitude lower than their abundances around a Sun-like star.

Dry deposition velocities of long-lived compounds, notably major volcanic carbon compounds including methane, carbon monoxide, and carbon dioxide, have significant effects on the atmospheric oxidation states. The specific choice of dry deposition velocities for emitted gases and their major photochemical byproducts in the atmosphere is critical to determine the atmospheric composition on terrestrial exoplanets.

Volcanic carbon compounds (i.e., \ce{CH4} and \ce{CO2}) are chemically long-lived and tend to be well mixed in terrestrial exoplanet atmospheres, whereas volcanic sulfur compounds (i.e., \ce{H2S} and \ce{SO2}) are short-lived. \ce{CH4} and \ce{CO2} have chemical lifetime longer than 10,000 years in all three benchmark atmospheres ranging from reducing to oxidizing, implying that a relatively small volcanic input can result in a high steady-state mixing ratio. The chemical lifetime \ce{CO}, another possible volcanic carbon compound, ranges from 0.1 to 700 years depending on the \ce{OH} abundance in the atmosphere.

We find abiotic \ce{O2} and \ce{O3}, photochemically produced from \ce{CO2} photolysis, build up in the 1-bar \ce{CO2} dominated atmosphere if volcanic emission rates of reducing gases (i.e., \ce{H2} and \ce{CH4}) is more than one order of magnitude lower than current Earth's volcanic rates. Abiotic \ce{O2} can be a false positive for detecting oxygenic photosynthesis, but the combination of \ce{O2}/\ce{O3} and reducing gases remains rigorous biosignature.

\subsection{Chemistry of \ce{H2}, \ce{N2}, and \ce{CO2} Dominated Atmospheres}

\begin{table}[htdp]
\caption{Atmospheric compositions of terrestrial exoplanet benchmark scenarios. For the surface emission in 1-bar \ce{H2}, \ce{N2}, and \ce{CO2} dominated atmospheres, mixing ratios of emitted gases, photochemical products and reactive agents are computed by the photochemistry model. }
\scriptsize
\begin{center}
\begin{tabular}{l|lll}
\hline
Scenario & \multicolumn{3}{l}{{Column Averaged Mixing Ratio }} \\
&  Emitted Gases & Photochemical Products\tablenotemark{a} & Reactive Agents\tablenotemark{b}  \\
\hline
Reducing 	
& \ce{CO2}: $8.9\times10^{-5}$ & \ce{CO}: $8.0\times10^{-6}$ & \ce{H}: $2.1\times10^{-9}$ \\
90\%\ce{H2}, 10\%\ce{N2}
& \ce{SO2}: $9.9\times10^{-12}$ & \ce{C2H6}: $4.7\times10^{-10}$ & \ce{OH}: $2.0\times10^{-14}$ \\
& \ce{CH4}: $5.9\times10^{-6}$ & \ce{S8}: $3.5\times10^{-10}$ & \ce{O}: $1.2\times10^{-11}$ \\
& \ce{H2S}: $9.1\times10^{-10}$ & \ce{CH2O}: $2.9\times10^{-10}$ & \ce{O(^1D)}: $2.2\times10^{-21}$ \\
&  & \ce{CH4O}: $5.6\times10^{-11}$ &  \\
\hline
Weakly oxidizing		
& \ce{CO2}: $1.3\times10^{-4}$ & \ce{CO}: $1.7\times10^{-7}$ & \ce{H}: $1.2\times10^{-9}$ \\
\ce{N2}
& \ce{H2}: $4.5\times10^{-4}$  &  \ce{C2H6}: $9.0\times10^{-9}$ & \ce{OH}: $9.3\times10^{-14}$ \\
& \ce{SO2}: $8.9\times10^{-12}$  & \ce{CH4O}: $1.4\times10^{-9}$ &  \ce{O}: $6.5\times10^{-10}$ \\
& \ce{CH4}: $3.1\times10^{-5}$ & \ce{O2}: $3.4\times10^{-10}$ & \ce{O(^1D)}: $1.8\times10^{-20}$ \\
& \ce{H2S}: $1.1\times10^{-14}$ & \ce{S8}: $3.0\times10^{-10}$ & \\
&   & \ce{CH2O}: $4.0\times10^{-11}$ & \\
&   & \ce{C2H2}: $1.5\times10^{-11}$ & \\
\hline
Highly oxidizing
& \ce{H2}: $1.0\times10^{-3}$ & \ce{CO}: $7.7\times10^{-3}$  & \ce{H}: $6.0\times10^{-11}$ \\
90\%\ce{CO2}, 10\%\ce{N2}
& \ce{SO2}: $1.6\times10^{-10}$  & \ce{O2}: $6.4\times10^{-7}$ & \ce{OH}: $7.8\times10^{-15}$ \\
& \ce{CH4}: $3.7\times10^{-5}$   & \ce{C2H6}: $6.1\times10^{-10}$ &  \ce{O}: $2.0\times10^{-8}$ \\
& \ce{H2S}: $1.4\times10^{-10}$  & \ce{H2O2}: $3.7\times10^{-10}$ & \ce{O(^1D)}:  $3.0\times10^{-18}$ \\
&   & \ce{H2SO4}: $5.0\times10^{-11}$ &  \\
&   & \ce{CH2O}: $2.5\times10^{-12}$ &  \\
\hline
\end{tabular}
\tablenotetext{a}{Species produced in the atmosphere from the emitted volcanic gases via photochemistry and subsequent series of chemical reactions, listed in the order of decreasing abundance. \ce{S8} and \ce{H2SO4} mixing ratios include both the gas phase and the condensed phase; whereas the condensed phase is found to contain more than 85\% of mass.}
\tablenotetext{b}{Common reactive agents in the atmosphere are \ce{H}, \ce{OH}, \ce{O}, \ce{O(^1D)}. We list the abundance of these gases in all three scenarios for as useful reference for the future assessment of chemical lifetime of trace gases.}
\end{center}
\normalsize
\label{AtmosScenario}
\end{table}

\subsubsection{\ce{H2} Dominated Atmospheres }

The main reactive agent in the \ce{H2} dominated atmosphere is atomic hydrogen (\ce{H}). The abundance of atomic hydrogen is five orders of magnitudes higher than that of hydroxyl radical (Figure \ref{Benchmark}). The source of both \ce{H} and \ce{OH} is water vapor photodissociation. In the \ce{H2} dominated atmosphere, most of the \ce{OH} molecules produced from water vapor photodissociation react with \ce{H2} to reform \ce{H2O} and produce \ce{H} through the reactions:
\reaction{H2O + hv -> H + OH  ,}
\reaction{OH + H2 -> H + H2O  .}
Therefore the abundance of atomic hydrogen is much higher than that of OH. We note that \ce{H} production via water photodissociation is much more efficient than the direct photodissociation of \ce{H2}, which requires radiation in wavelengths less than 85 nm. As water vapor is the primary source of \ce{H} and \ce{OH} in anoxic atmospheres, the amounts of \ce{H} and \ce{OH} depends on the mixing ratio of water vapor above the cold trap, which is in turn sensitively controlled by the cold trap temperature. Water vapor mixing ratio spans 3 orders of magnitudes for the cold trap temperature ranging in 160 - 200 K; consequently, the number densities of \ce{H} and \ce{OH} in the atmosphere can easily vary by one order of magnitude, depending on the cold trap temperature. The removal of atomic hydrogen is mainly by recombination to \ce{H2}, which can be more efficient with the presence of \ce{CO}, via
\reaction{H + CO + M -> CHO + M , \label{CHO1}}
\reaction{H + CHO -> CO + H2 . \label{CHO2}}

As a result of low \ce{OH} abundance in the \ce{H2} dominated atmosphere, both \ce{CH4} and \ce{CO} are long-lived and therefore well mixed. \ce{CH4} emitted from the surface is slowly oxidized in the atmosphere into methanol, which gives methane a chemical lifetime of $8\times10^4$ years. \ce{CO} is produced by \ce{CO2} photodissociation in our benchmark model; it is emitted by volcanoes on Earth at a much lower rate than \ce{CO2}, but it can presumably be the main carbon-bearing gas produced by volcanoes if the upper mantle is more reduced than the current Earth's (e.g., Holland 1984). We find that \ce{CO} is long-lived in the \ce{H2} atmosphere as well.

The lack of efficient atmospheric sink of \ce{CH4} and \ce{CO} in the \ce{H2} atmosphere implies that surface sink, if any, is the major sink for these two carbon compounds. \ce{CH4} and \ce{CO} have zero or very small dry deposition velocities on an abiotic planet, so they can build up to significant amounts in the \ce{H2} atmosphere (Figure \ref{Benchmark}). If a nonzero deposition velocity is adopted for \ce{CH4} and \ce{CO}, their steady-state mixing ratios will be much lower than the benchmark model. For example, using a deposition velocity for \ce{CH4} standard on Earth ($\sim0.01$ cm s$^{-1}$) results in a mixing ratio of less than 1 ppb for \ce{CH4}, compared to a mixing ratio of 6 ppm in the benchmark model. The dry deposition velocity is indeed the controlling factor for the steady-state abundance of the long-lived carbon compounds. In comparison, the emitted sulfur compounds (\ce{H2S} and \ce{SO2}) are short-lived. In the \ce{H2} atmosphere, sulfur emission from the surface is readily converted to elemental sulfur aerosols (\ce{S8}) in the atmosphere.

Interestingly, \ce{CO2} is fairly long-lived and well mixed in the \ce{H2} atmosphere, meaning that the reduction of \ce{CO2} in the \ce{H2} dominated atmosphere is not efficient (Figure \ref{Benchmark}). Only 1/3 of emitted \ce{CO2} is reduced in the atmosphere, and the rest is deposited to the surface by dry and wet deposition. It is the balance between surface emission and surface deposition that sets the steady-state mixing ratio of \ce{CO2}.

\subsubsection{\ce{N2} Dominated Atmospheres }

Both reducing radicals (i.e., \ce{H}) and oxidizing radicals (i.e., \ce{O} and \ce{OH}) are relatively abundant in the \ce{N2} atmosphere compared with the \ce{H2} atmosphere (Figure \ref{Benchmark}). Like in \ce{H2} dominated atmospheres, \ce{H} abundance is orders of magnitude larger than \ce{OH} abundance, because most of \ce{OH} molecules from water photolysis react with \ce{H2} to reform \ce{H2O} and produce \ce{H}. The molecular hydrogen that consumes \ce{OH} and boosts \ce{H} is emitted volcanically, so a general \ce{N2} atmosphere can be more oxidizing (i.e., having lower \ce{H} and higher \ce{OH}) if there is a lower volcanic \ce{H2} emission and mixing ratio than the benchmark model.

An important feature of the \ce{N2} dominated atmosphere is that both \ce{H} and \ce{OH} are relatively abundant near the surface. Comparing the \ce{N2} dominated atmosphere with the \ce{H2} dominated atmosphere (Figure \ref{Benchmark}), we find that both the \ce{OH} and the \ce{H} number densities are higher near the surface due to the lower \ce{H2} number density. The relatively high \ce{OH} number density leads to relatively fast removal of \ce{CO} by
\reaction{CO + OH -> CO2 + H , \label{COOH}}
as shown in Figure \ref{Benchmark}. With a low \ce{CO} abundance the recombination of \ce{H} via reactions (\ref{CHO1}-\ref{CHO2}) is inefficient. Counter-intuitively, a high \ce{OH} number density helps preserve \ce{H} in this specific case. This example shows the complexity and the nonlinearity of an atmospheric chemical network. The feature of simultaneous high \ce{OH} and \ce{H} abundances near the surface is sensitive to the specification of surface hydrogen emission and eddy diffusion coefficients (see section \ref{ParaPhysics} for relevant rationale).

The chemical lifetimes of \ce{CH4} and \ce{CO} mainly depend on the amount of \ce{OH}. In the \ce{N2} atmosphere, \ce{CH4} is well mixed because its chemical lifetime is long. \ce{CH4} is photodissociated and oxidized slowly in the atmosphere into methanol (\ce{CH4O}) with a chemical lifetime of $\sim6\times10^4$ years. The photolysis of methane is a secondary source of atomic hydrogen, which concentrates at the pressure level of 10 Pa. Interestingly, methane photolysis causes the apparent trough of \ce{O2} mixing ratio profile at $\sim10$ Pa. It is the shielding of the UV radiation that dissociates methane by methane itself that determines this pressure level (see Appendix D for an analytical formula for assessing the pressure level at which the photolysis of a certain gas is important).

\ce{CO2} is actively photodissociated into \ce{CO} and \ce{O} in the upper atmosphere, but most of the \ce{CO} produced is efficiently converted back to \ce{CO2} by reacting with \ce{OH} (reaction \ref{COOH}). In equilibrium the net chemical removal of \ce{CO2} is minimal, so the \ce{CO2} mixing ratio is set by the balance between emission and deposition. The steady-state amount of \ce{CO2} can be directly estimated by the mass balance between emission and deposition, viz.,
\begin{equation}
F = f_1 = \frac{n_1}{N_1} = \frac{\Phi_{1/2}}{V_{\rm DEP}N_1} , \label{SimpleBoundary}
\end{equation}
where the overall mixing ratio ($F$) is the same as the near-surface mixing ratio ($f_1$), and $\Phi_{1/2}$ is the surface emission rate. For \ce{CO2} in the \ce{N2} dominated atmosphere, using equation (\ref{SimpleBoundary}), with $V_{\rm DEP}(\ce{CO2})=1\times10^{-4}$ cm s$^{-1}$, we find the steady-state mixing ratio is $1.3\times10^{-4}$, consistent with the full photochemical model (see Table \ref{AtmosScenario} and Figure \ref{Benchmark}). The steady-state mixing ratio of a long-lived volcanic gas that is primarily removed by surface deposition is inversely proportional to its dry deposition velocity.

Sulfur-bearing gases emitted from the surface are effectively converted into elemental sulfur and sulfuric acids. Elemental sulfur, mainly in the condensed phase (i.e., aerosols), is the major sulfur-bearing species in the steady state because of relatively high \ce{H2} mixing ratio in the benchmark model. In separate numerical simulations we find that sulfuric acid aerosols may outnumber elemental sulfur aerosols when the \ce{H2} emission is reduced by more than 1 order of magnitude and the atmosphere is more oxidizing than the benchmark model.

\subsubsection{\ce{CO2} Dominated Atmospheres }

\ce{CO2} photodissociation produces \ce{CO}, \ce{O}, and \ce{O2}. Atomic oxygen is the most abundant reactive radicals in the \ce{CO2} dominated atmosphere (Figure \ref{Benchmark}), because \ce{H} is readily removed by \ce{O2} and \ce{OH} is readily removed by \ce{CO}. As a result of low \ce{H} and \ce{OH}, \ce{CO} is long-lived in the atmosphere and can build up to very high mixing ratios in the atmosphere depending on its dry deposition velocity; and \ce{CH4} emitted from the surface is also long-lived in the atmosphere with a chemical lifetime of $\sim6\times10^4$ years. The steady-state abundance of \ce{CO} and \ce{CH4} is therefore controlled by their dry deposition velocities. Most of the emitted \ce{SO2} is deposited to the surface, because there are very few \ce{OH} or \ce{O} radicals near the surface. A small fraction of the \ce{SO2} is transported upwards and converted into sulfuric acid aerosols in the radiative layer.

Based on our benchmark model of \ce{CO2} dominated atmospheres, we now revisit whether photochemically produced \ce{O2} can cause false positive for detecting oxygenic photosynthesis. Oxygen and ozone are the most studied biosignature gases for terrestrial exoplanet characterization (e.g., Owen 1980; Angel 1986; L\'eger et al. 1993, 1996; Beichman et al. 1999). One of the main concerns of using \ce{O2}/\ce{O3} as biosignature gases is that \ce{O2} may be produced abiotically from photodissociation of \ce{CO2}. A number of authors have studied the abiotic production of oxygen in terrestrial atmospheres, either for understanding prebiotic Earth's atmosphere (e.g., Walker 1977; Kasting et al. 1979; Kasting \& Catling 2003), or for assessing whether abiotic oxygen can be a false positive for detecting photosynthesis on habitable exoplanets (Selsis et al. 2002; Segura et al. 2007). It has been proposed that abiotic oxygen from \ce{CO2} photodissociation is not likely to build up in the atmosphere on an planet having active hydrological cycle (Segura et al. 2007).

\begin{figure}[h]
\begin{center}
 \includegraphics[width=0.8\textwidth]{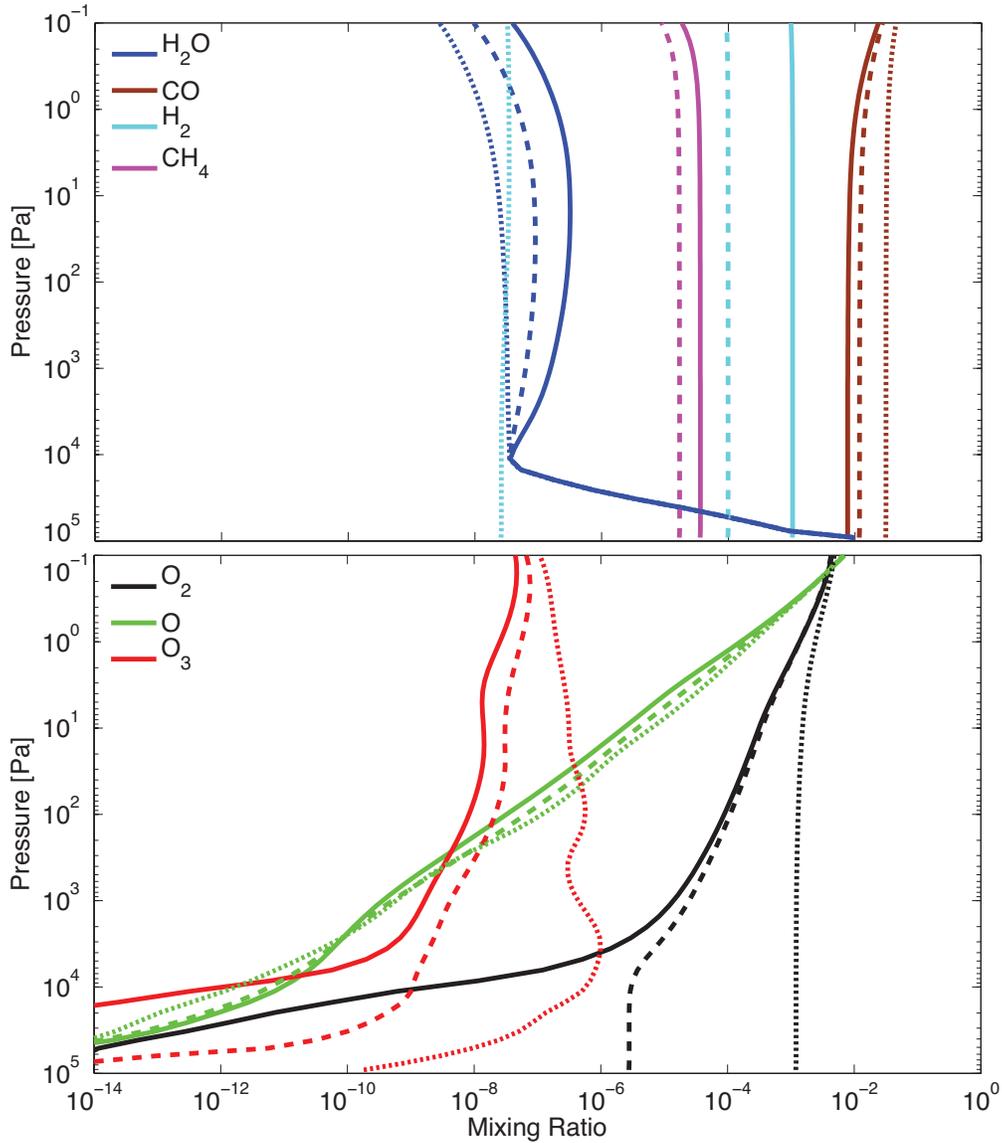}
 \caption{
 Effects of the surface volcanic emission for \ce{CO2}-dominated atmospheres on rocky exoplanets. The upper panel shows mixing ratios of \ce{H2O}, \ce{CO}, \ce{H2}, and \ce{CH4}, and the lower panel shows mixing ratios of \ce{O2}, \ce{O}, and \ce{O3}.
 The solid lines show the chemical composition of the benchmark scenario whose parameters are tabulated in Table \ref{AtmosPara}. In particular the emission rate of \ce{H2} is $3\times10^{10}$ cm$^{-2}$ s$^{-1}$.  The dashed lines show the chemical composition of the same scenario, but with an \ce{H2} emission rate of $3\times10^{9}$ cm$^{-2}$ s$^{-1}$; and the dotted lines show the chemical composition for zero emission of \ce{H2} and \ce{CH4}.
 We see a dramatic increase of \ce{O2} and \ce{O3} mixing ratios as a result of a decrease of the surface emission of reduced gases.
}
\label{CO2_compare}
\end{center}
\end{figure}

We find that the steady-state number density of \ce{O2} and \ce{O3} in the \ce{CO2} dominated atmosphere is mainly controlled by the surface emission of reducing gases such as \ce{H2} and \ce{CH4}, and without surface emission of reducing gas photochemically produced \ce{O2} can build up in a 1-bar \ce{CO2} dominated atmosphere. In addition to the benchmark model, we have simulated \ce{CO2} dominated atmospheres with relatively low and zero emission rates of \ce{H2} and \ce{CH4} (Table \ref{redoxbalance} and Figure \ref{CO2_compare}). We see that the \ce{O2} mixing ratio near the surface increases dramatically in 1-bar \ce{CO2} atmospheres when the emission of reducing gases decreases. \ce{O2} is virtually nonexistent at the surface for the Earth-like emission rates of \ce{H2} and \ce{CH4}; but \ce{O2} mixing ratio can be as high as $10^{-3}$ if no \ce{H2} or \ce{CH4} is emitted (Figure \ref{CO2_compare}). In particular, if no \ce{H2} or \ce{CH4} is emitted, the \ce{O3} column integrated number density can reach one third of the present-day Earth's atmospheric levels (Table \ref{redoxbalance}), which constitutes a potential false positive.

In the 1-bar \ce{CO2} dominated atmosphere \ce{O3} can potentially build up to a false-positive level even on a planet with active hydrological cycle. Segura et al. (2007) have based their conclusion on simulations of 20\% \ce{CO2} 1-bar atmospheres with and without emission of \ce{H2} and \ce{CH4} and simulations of 2-bar \ce{CO2} atmospheres with emission of \ce{H2} and \ce{CH4}. We have been able to reproduce all results of Segura et al. (2007) quantitatively to within a factor of two. Where we differ from Segura et al. (2007) is that we successfully simulated high \ce{CO2} 1-bar atmospheres with minimal volcanic reducing gas emission (Figure \ref{CO2_compare}). This is a parameter space that Segura et al. (2007) did not cover, but we find that this is the parameter space for high abiotic \ce{O2}.

\begin{table}[htdp]
\caption{Mixing ratios of \ce{O2} and \ce{O3} and redox budget for \ce{CO2}-dominated atmospheres on rocky exoplanets having different surface emission of reducing gases. The redox number for each species is defined according to equation (\ref{redox}). In the redox balance, all values have unit of molecule cm$^{-2}$ s$^{-1}$, and defined as positive for hydrogen flux into the atmosphere. The redox budget for our atmosphere models is balanced, meaning that the atmosphere is not becoming more oxidized or reduced.}
\begin{center}
\begin{tabular}{llll}
\hline
\hline
Chemical & \multicolumn{3}{l}{\ce{CO2}-dominated atmospheres} \\
species & $\Phi_{1/2}$(\ce{H2}) =  $3\times10^{10}$ cm s$^{-1}$ & $\Phi_{1/2}$(\ce{H2}) =  $3\times10^{9}$ cm s$^{-1}$ & No \ce{H2} Emission  \\
\hline
\hline
\multicolumn{4}{l}{{\it Column-averaged mixing ratio}} \\
\ce{O2} & 6.4E-7  & 3.8E-6  & 1.3E-3 \\
\ce{O3} & 7.0E-11 & 3.7E-10 & 1.3E-7 \\
\hline
\hline
\multicolumn{4}{l}{{\it Redox balance}} \\
\multicolumn{4}{l}{Atmospheric escape} \\
\ce{H} & -7.0E+8 & -2.4E+8 & -1.2E+6 \\
\ce{H2} & -5.9E+10 & -5.8E+9 & -2.0E+6 \\
\hline
\multicolumn{4}{l}{Surface emission} \\
\ce{H2} & 6.0E+10 & 6.0E+9 & 0  \\
\ce{CH4} & 2.4E+9 & 2.4E+9 & 0\\
\ce{H2S} & 1.8E+9 & 1.8E+9 & 1.8E+9 \\
\hline
\multicolumn{4}{l}{Dry and wet deposition} \\
\ce{O3} & 0 & 6.3E+4 & 1.1E+10 \\
\ce{HO2} & 3.1E+4 & 7.8E+7 & 1.2E+8 \\
\ce{H2O2} & 1.7E+5 & 1.5E+9 & 2.7E+9 \\
\ce{CO} & -3.7E+9 & -5.6E+9 & -1.5E+10 \\
\ce{CH2O} & -1.3E+6 & -7.2E+4  & 0 \\
\ce{Organic Haze} & -2.0E+6 & -8.1E+1 & 0 \\
\ce{H2S} & -4.9E+8 & -2.5E+8 & -3.2E+8 \\
\ce{H2SO4} & 1.2E+8 & 6.3E+7 & 2.9E+7\\
\hline
Balance & 4.9E+3 & -9.5E+2 & -1.3E+3 \\
\hline
\hline
\end{tabular}
\end{center}
\label{redoxbalance}
\end{table}

\subsection{Rationale of Model Parameters}

\label{ParaPhysics}

We here provide rationale for our specification of the atmospheric temperature profile, the eddy diffusion coefficients, and the dry deposition velocities.

First, the surface temperatures for the three scenarios are assumed to be 288 K, and the semi-major axis of the planet is adjusted according to appropriate amounts of greenhouse effect. The semi-major axis around a Sun-like star for \ce{H2}, \ce{N2}, and \ce{CO2} dominated atmospheres is found to be 1.6 AU, 1.0 AU, and 1.3 AU. We have compared the planetary thermal emission flux and the incidence stellar flux to determine the semi-major axis, a similar procedure as Kasting et al. (1993) taking into account \ce{CO2}, \ce{H2O}, and \ce{CH4} absorption, \ce{H2} collision-induced absorption, and 50\% cloud coverage and a Bond albedo of 30\%. The temperature profiles are assumed to follow appropriate dry adiabatic lapse rate (i.e., the convective layer) until 160 K (\ce{H2} atmosphere), 200 K (\ce{N2} atmosphere), and 175 K (\ce{CO2} atmosphere) and to be constant above (i.e., the radiative layer). We simulate the atmosphere up to the altitude of about 10 scale heights, or the pressure level of 0.1 Pa. The adopted temperature profiles are consistent with significant greenhouse effects in the convective layer and no additional heating above the convective layer for habitable exoplanets. The results discussed above do not change significantly if these temperature profiles are changed by several tens of K.

Second, we have used eddy diffusion coefficients empirically determined on Earth and scaled the values to account for different mean molecular masses. The eddy diffusion coefficient at a certain pressure level is assumed to be that of current Earth's atmosphere at the same pressure level (see Figure \ref{Transport}), scaled by 6.3, 1.0, and  0.68 for the \ce{H2}, \ce{N2}, and \ce{CO2} dominated atmospheres, respectively to account for different dominant molecules. We have roughly scaled the eddy diffusion coefficient assuming $K\propto H_0$ where $H_0$ is the atmospheric scale height. The reasoning of the scaling is as follows. According to the mixing length theory, $K\propto lw$, where $l$ is the typical mixing length and $w$ is the mean vertical velocity. The mixing length is a fraction of the pressure scale height (e.g., Smith 1998). The mean vertical velocity is related to the vertical convective energy flux, as $F \propto pw$, where $p$ is the pressure (e.g., Lindzen 1990). For a certain planet $F$ should have the same order of magnitude for different atmosphere compositions. As we apply the scaling from pressure surface to pressure surface, we have roughly $K \propto H_0$.  The scaling is an approximation and we only intend to provide a consistent description of eddy diffusion for atmospheres with very different mean molecular mass. The eddy diffusion coefficients for the three scenarios are also consistent with their temperature profiles, featuring minima near the tropopause. The general results discussed above are not sensitive to the variation of eddy diffusion coefficients by an order of magnitude.


Third, the typical deposition velocity on Earth is sometimes not directly applicable for terrestrial exoplanets, because the major surface sink of a number of gases (notably, \ce{H2}, \ce{CH4}, \ce{CO}) on Earth is actually microorganisms. In this paper we focus on the scenarios assuming no biotic contribution in neither surface emission nor surface deposition. For \ce{H2} and \ce{CH4}, a sensible dry deposition velocity without biotic surface sink is zero (J. Kasting, 2012, private communication). For \ce{CO2}, the dry deposition velocity should match with the timescale of weathering and carbonate formation, which is 10,000 years (Archer et al. 2010). For \ce{CO}, the rate limiting step for converting \ce{CO} into bicarbonate has been proposed to the hydration of \ce{CO} in the ocean, which corresponds to a deposition velocity of $1.0\times10^{-9}\sim10^{-8}$ cm~s$^{-1}$ (Kharecha et al. 2005). The adopted or assumed values of dry deposition velocities are tabulated in Table \ref{AtmosPara}.

For completeness, we now comment on our assumptions for nitrogen chemistry and organic haze. We do not track the bulk nitrogen cycle, instead assuming 10\% \ce{N2} for \ce{H2} and \ce{CO2} dominated atmospheres. Our model does not treat abiotic nitrogen fixation by lightning (see Kasting \& Walker 1981; Zahnle 1986; Tian et al. 2011 for specific analysis), so \ce{N2} is considered as inert in our models. We include the formation of elemental sulfur aerosols and sulfuric acid aerosols in our models; but we do not include the less-understood hydrocarbon chemical network for hydrocarbon molecules that have more than 2 carbon atoms or the formation of hydrocarbon haze. Organic haze may be formed in anoxic atmospheres based on methane photolysis and hydrocarbon polymerization. The number density of hydrocarbons that have more than 4 carbon atoms (usually considered to be condensable) is $\sim4$ orders of magnitude smaller than the number density of \ce{C2H6} at the steady state (e.g. Allen et al. 1980; Yung et al. 1984; Pavlov et al. 2001). To account for the loss of carbon due to possible organic haze formation and deposition, we apply an {\it ad hoc} dry deposition velocity of $1.0\times10^{-5}$ cm s$^{-1}$ for \ce{C2H6}. This small velocity results from a scaling based on typical sub-micron particle deposition velocity ($\sim0.2$ cm s$^{-1}$, Sehmel 1980) and the number density ratio of \ce{C4} hydrocarbons and \ce{C2H6} (e.g., Yung et al. 1984). An accurate treatment for organic haze is not the purpose of this paper, because the efficiency of haze formation depends on oxidation states of atmospheres, stellar UV radiation, and a number of less-understood reaction rates. Nonetheless, for all simulated scenarios, the carbon loss due to haze formation is less than 1\% of the methane emission flux; therefore we do not expect our simplification of organic haze formation would impact our results. We will fully explore the formation of organic haze and its implication on atmospheric chemistry in a separate paper (Hu et al., in prep.).

\section{Summary}

We have developed a comprehensive photochemistry model for the study of terrestrial exoplanet atmospheres. The photochemistry model solves the one-dimensional chemical-transport equation for 111 O, H, C, N, S species including \ce{S8} and \ce{H2SO4} aerosols. The output is the steady state of molecular mixing ratios in which concentrations of all species at all altitudes do not vary. In order to find the steady-state solution from arbitrary initial conditions for a wide variety of atmospheric compositions, required for the study of exoplanets, we have designed a numerical scheme that allows the selection of chemical species to be treated in fast or slow reactions automatically. The steady-state solution depends on a pool of input parameters, among which the major chemical species, surface emission, deposition velocities of long-lived species, and ultraviolet radiation are found to be critical. We validate the photochemistry model by simulating the atmospheric composition of current Earth and Mars.

Based on the photochemistry model, we have investigated the main chemistry processes and lifetimes of key spectrally active species for rocky exoplanet atmospheres by simulating benchmark cases of atmospheres having redox states ranging from reducing to oxidizing. We find that atomic hydrogen is a more abundant reactive radical than hydroxyl radical in anoxic atmospheres, and therefore reactions with atomic hydrogen are likely to be an important removal pathway for spectrally important trace gases. The source of \ce{H} and \ce{OH} is water vapor photolysis in anoxic atmospheres, and the abundance of \ce{H} in the atmosphere is always larger than the amount of \ce{OH} because \ce{OH} can react with \ce{H2} or \ce{CO} to produce \ce{H}. In addition to atomic hydrogen, in weakly oxidizing \ce{N2} atmospheres, \ce{OH}, despite its lower abundance than \ce{H}, is important in removing \ce{CH4} and \ce{CO}. In highly oxidizing \ce{CO2} atmospheres, atomic oxygen is the most abundant reactive species.

As a general observation we find that volcanic carbon compounds are long-lived and volcanic sulfur compounds are short-lived. In particular, due to the scarcity of \ce{OH} in anoxic atmospheres, methane is always long-lived having chemical lifetime longer than 10,000 years. We also find that the reduction of \ce{CO2} to \ce{CO} and formaldehyde is minimal in \ce{N2} atmospheres and limited in \ce{H2} dominated atmospheres. In contrast to carbon species, volcanic sulfur compounds (i.e., \ce{H2S} and \ce{SO2}) are readily converted into either elemental sulfur or sulfuric acid aerosols in atmospheres from reducing to oxidizing. We will discuss in detail the sulfur chemistry in an accompanying paper (Hu et al. 2012).

The photochemistry is critical for prospecting the possible atmospheric composition that will eventually be characterized by a TPF-like mission. We have shown that volcanic carbon compounds including \ce{CH4} and \ce{CO2} are likely to be abundant in terrestrial exoplanet atmospheres; and in the accompany paper we will show that an enhanced volcanic activity leads to formation of optically thick sulfur or sulfate aerosols. As for biosignatures, we here have shown that photochemically produced \ce{O2} and \ce{O3} can be a potential false positive biosignature in thick \ce{CO2} atmospheres. We also find that oxygen and ozone can only build up without \ce{H2} and \ce{CH4} emission; so we confirm that simultaneous detection of ozone and methane remains a rigorous biosignature. More generally, the three benchmark models presented in this paper can serve as the standard atmospheres for reducing, weakly oxidizing, and highly oxidizing atmospheres on habitable exoplanets for assessing chemical lifetime of potential biosignature gases.

\acknowledgments

We thank James Kasting for helpful suggestions about the photochemical model. We thank Kerry Emanuel for enlightening discussion about modeling the Earth's hydrological cycle. We thank Linda Elkins-Tanton for helpful suggestions on the mantle degassing. We thank Susan Solomon for discussions about \ce{CO2} cycle on terrestrial exoplanets. We thank the anonymous referee for the improvement of the manuscript. RH is supported by the NASA Earth and Space Science Fellowship (NESSF/NNX11AP47H).

\appendix

\section{Formulation of Vertical Diffusion Flux}

\label{A_VD}

The vertical diffusion flux can be derived rigorously from the general diffusion equation for a minor constituent in a heterogenous atmosphere that
\begin{equation}
\Phi=-K\bigg[\frac{\partial n}{\partial z}+n\bigg(\frac{1}{H_0}+\frac{1}{T}\frac{dT}{dz}\bigg)\bigg]
-D\bigg[\frac{\partial n}{\partial z}+n\bigg(\frac{1}{H}+\frac{1+\alpha_{\rm T}}{T}\frac{dT}{dz}\bigg)\bigg],\label{eq_diffusion_g}
\end{equation}
in which the first term is the eddy diffusion flux ($\Phi_{K}$) and the second term is the molecular diffusion flux ($\Phi_{D}$). Note that $n=Nf$, and then the eddy diffusion flux can be simplified as
\begin{eqnarray}
\Phi_{K} & = & -K\bigg[\frac{\partial n}{\partial z}+n\bigg(\frac{1}{H_0}+\frac{1}{T}\frac{dT}{dz}\bigg)\bigg],\nonumber\\
& = &  -K\bigg[N\frac{\partial f}{\partial z}+fN\bigg(\frac{1}{N}\frac{\partial N}{\partial z}+\frac{1}{H_0}+\frac{1}{T}\frac{dT}{dz}\bigg)\bigg], \nonumber\\
& = &  -K\bigg[N\frac{\partial f}{\partial z}+fN\bigg(\frac{1}{N}\frac{\partial N}{\partial z}-\frac{1}{P}\frac{\partial P}{\partial z}+\frac{1}{T}\frac{dT}{dz}\bigg)\bigg], \nonumber\\
& = & -KN\frac{\partial f}{\partial z},
\end{eqnarray}
in which we have used the definition of atmospheric scale height and the ideal gas law.

The fundamental difference between the molecular diffusion and the eddy diffusion is in the scale height term. The molecular diffusion depends on the specific scale height for the molecule, whereas the total atmospheric pressure falls according to the mean scale height. The molecular diffusion term can be similarly simplified as
\begin{eqnarray}
\Phi_{D} & = & -D\bigg[\frac{\partial n}{\partial z}+n\bigg(\frac{1}{H}+\frac{1+\alpha_{\rm T}}{T}\frac{dT}{dz}\bigg)\bigg] ,\nonumber\\
& = & -D\bigg[\frac{\partial n}{\partial z}+n\bigg(\frac{1}{H_0} + \frac{1}{T}\frac{dT}{dz}\bigg) - n\bigg( \frac{1}{H_0} - \frac{1}{H} -\frac{\alpha_{\rm T}}{T}\frac{dT}{dz}\bigg)\bigg] ,\nonumber\\
& = & -DN\frac{\partial f}{\partial z} + Dn\bigg(\frac{1}{H_0}-\frac{1}{H}-\frac{\alpha_{\rm T}}{T}\frac{dT}{dz}\bigg) ,
\end{eqnarray}
which yields equation (\ref{eq_diffusion}).

\section{Mean Stellar Zenith Angle for 1-Dimensional Photochemistry Model}

\label{A_Zenith}

All 1-dimensional photochemistry models need to assume a zenith angle ($\theta_0$) for the incoming stellar radiation as a global average. Various values have been adopted in previous photochemistry models, for example $\theta_0=50^{\circ}$ (Zahnle et al. 2006), $\theta_0=57.3^{\circ}$ (Zahnle et al. 2008), and $\theta_0=48^{\circ}$ (Moses et al. 2011). We find that all these assumptions are plausible and provide justification as follows.

\begin{figure}[h]
\begin{center}
 \includegraphics[width=0.5\textwidth]{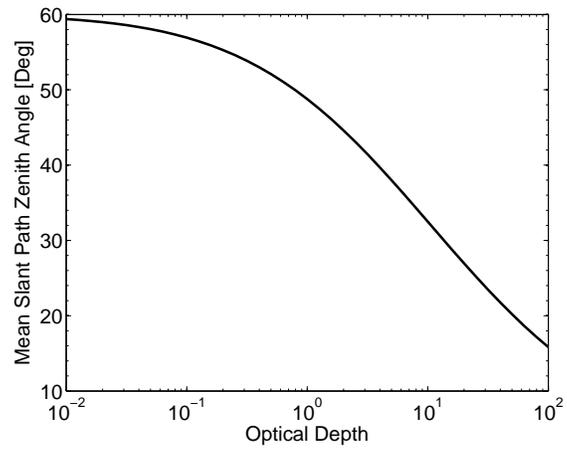}
 \caption{Mean zenith angle of stellar radiation for photochemistry models as a function of optical depth.}
 \label{Slant}
  \end{center}
\end{figure}

At different location of star-facing hemisphere of the planet, the local stellar zenith angle is $\mu\equiv\cos\theta=\cos\psi\cos\phi$ where $\psi$ and $\phi$ are the local latitude and longitude, respectively. A dayside disk average should be weighted by the radiation intensity at certain optical depth $\tau$, viz.
\begin{equation}
\exp(-\tau/\mu_0) = \frac{1}{\pi}\int_{-\pi/2}^{\pi/2}\int_{-\pi/2}^{\pi/2}
\exp(-\tau/\cos\psi\cos\phi)\cos^2\psi \cos\phi d\psi d\phi  . \label{eq_zenith}
\end{equation}
Equation (\ref{eq_zenith}) is well-defined for any particular level of optical depth, and can be solved numerically. The relationship between the average zenith angle and the optical depth of concern is illustrated in Figure \ref{Slant}.

We see in Figure \ref{Slant} that the appropriate mean zenith angle depends on the optical depth of concern. We find that optical depth between 0.1 and 1.0 corresponds to a mean zenith angle between $57^{\circ}$ and $48^{\circ}$. In the extreme of zero optical depth, the appropriate mean zenith angle is $60^{\circ}$. In general, it is appropriate to assume the mean zenith angle to be $48^{\circ}\sim60^{\circ}$ for the application of 1-dimensional photochemistry model.

\section{Deposition Velocities}

\label{A_DD}

In the photochemical model we calculate the number density in the bottom layer of the atmosphere. The interaction between the bottom layer and the surface consists of two steps: first, the molecular transport across a thin stagnant layer of air adjacent to the surface, called the {\it quasi-laminar sublayer}; second, the uptake at the surface. Each step provides a resistance to the overall dry deposition, viz.
\begin{equation}
V_{\rm DEP}^{-1} = r_b + r_c  ,
\end{equation}
in which $r_b$ is the quasi-laminar resistance and $r_c$ is the surface resistance. The quasi-laminar resistance is
\begin{equation}
r_b = \frac{5\ {\rm Sc}^{2/3}}{u_*} ,
\end{equation}
in which Sc is the dimensionless Schmidt number defined as the ratio between the kinetic viscosity of air and the molecular diffusivity of the molecule considered, and $u_*$ is the friction velocity (Seinfeld \& Pandis 2006). The friction velocity depends on the wind speed adjacent to the surface and the roughness of the surface. In the current model, the friction velocity $u_*$ should be treated as a free parameter varying from 0.1 to 1 m s$^{-1}$.

The surface resistance is more complicated and largely depends on the property of the surface and the solubility of the molecule. For example, we consider two general types of surfaces, ocean and land. For the land we envisage desert-like surface and do not consider the complications of foliage.

For the deposition to the ocean, the surface resistance is
\begin{equation}
r_c = \frac{1}{k_G}+\frac{1}{k_LH} ,
\end{equation}
in which $k_G$ is the gas phase mass transfer coefficient, $k_L$ is the liquid phase mass transfer coefficient and $H$ is the dimensionless Henry's law constant (Seinfeld \& Pandis 2006). These parameters depend on the wind speed (or the friction velocity), and the Schmidt number of the molecule in sea water.

The surface resistance to the land is even more complicated and depends on the properties of land, for example, surface morphology, roughness, vegetation, canopy, etc. There have been tremendous efforts to measure and parameterize the surface resistance of different molecules on different types of land on Earth. For example, the surface resistance to a featureless desert can be expressed as
\begin{equation}
r_c = \bigg(\frac{10^{-5}H}{r_S}+\frac{f_0}{r_O}\bigg)^{-1} ,
\end{equation}
in which $H$ is in the unit of M atm$^{-1}$, $f_0$ is a normalized (0 to 1) reactivity factor, $r_S=1000$ s m$^{-1}$ and $r_O=400$ s m$^{-1}$ (Seinfeld \& Pandis, 2006).

\begin{figure}[h]
\begin{center}
 \includegraphics[width=0.5\textwidth]{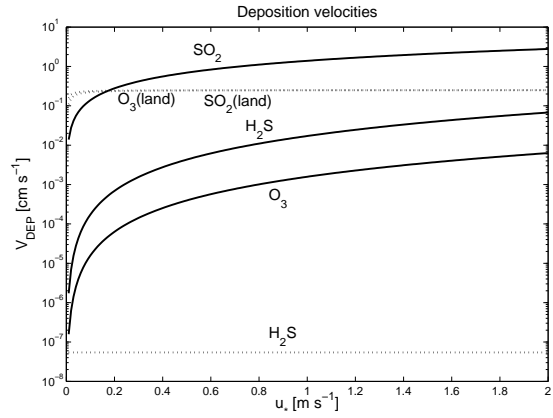}
 \caption{Deposition velocities to the ocean (solid lines) and the land (dotted lines) as a function of the friction velocity $u_*$. The calculation is for the terrestrial atmosphere at $T=273.15$ K, $P=1$ atm. The ocean is assumed to have a pH of about 6.5. The land is assumed to be featureless desert. }
 \label{depo}
  \end{center}
\end{figure}

We present the deposition velocities of slightly soluble gases (such as \ce{H2S}), highly soluble gases (such as \ce{SO2}), not soluble but reactive gases (such as \ce{O3}) to the ocean and the land in Figure \ref{depo}. The deposition velocity to the ocean critically depends on the solubility of the gas. For poorly soluble gas, the deposition velocity to the ocean is negligible. For the highly soluble gas, the deposition velocity is limited by the friction velocity, or the wind speed. For the reactive gas, the deposition velocity can be also very high. In particular, the deposition velocity of \ce{H2S} to the land the negligible, and in most cases, the deposition velocity of \ce{H2S} to the ocean is smaller than 0.1 cm s$^{-1}$. The estimate of dry deposition velocity here only considers the surface uptake, but not the eventual loss of a gas at or beneath the surface. The effective dry deposition velocity might be smaller than what we estimate if no effective loss mechanism is available at the surface and the surface is saturated (e.g., Kharecha et al. 2005).

\section{Photochemical Stability of Atmospheric Trace Gases}

We here present an analytical treatment of the photochemical stability of radiative trace gases. The goal of the analytical treatment is to obtain a useful formula that computes the critical pressure level of any radiative trace gas above which the photodissociation of the gas is important. The key idea is that the photolysis rate at any certain altitude (denoted as $z^*$) is proportional to the radiation flux in the dissociating wavelengths at this altitude, which is attenuated by absorptions above $z^*$. The problem can be significantly simplified, if the absorption is mainly due to the gas itself. This is an analog to the concept of the Chapman layer in the ionosphere (e.g., Banks \& Kockarts 1973).

In the following we first derive the analytical formula for the case of self-shielding and then extend the formula to the generic situation.

We intend to relate the ultraviolet optical depth to the pressure level. The ultraviolet optical depth in the dissociating wavelength range of a specified gas is
\begin{equation}
\tau_{\rm UV} = \frac{\sigma_{\rm UV}}{\mu_0}\int_{z^*}^{\infty}n(z)\ dz  ,
\end{equation}
where $\sigma_{\rm UV}$ is the characteristic cross section in the wavelengths that lead to photolysis. Under hydrostatic conditions,
\begin{equation}
P_{\rm atm} = M_{\rm atm}g\int_{z^*}^{\infty}N(z)\ dz ,
\end{equation}
where $P_{\rm atm}$ is the atmospheric pressure at the altitude of $z^*$ and $M_{\rm atm}$ is the mean molecular mass. We can define the column averaged mixing ratio as
\begin{equation}
F = \frac{\int_{z^*}^{\infty}n(z)\ dz}{\int_{z^*}^{\infty}N(z)\ dz}  ,
\end{equation}
and then the UV optical depth at $z^*$ can be related to the atmospheric pressure as
\begin{equation}
\tau_{\rm UV} = \frac{\sigma_{\rm UV} F P_{\rm atm}}{\mu_0g M_{\rm atm}}  .
\end{equation}
The molecule is subject to rapid photodissociation if the UV optical depth is smaller than unity, or $\tau_{\rm UV}<1$, so that the critical pressure level above which photolysis is important is
\begin{equation}
P^* = \frac{\mu_0g M_{\rm atm}}{\sigma_{\rm UV} F}  . \label{PhotoStability}
\end{equation}
The equation (\ref{PhotoStability}) can be readily extended to include other gases in the atmosphere may provide additional shielding, viz.,
\begin{equation}
P^* = \frac{\mu_0g M_{\rm atm}}{\sigma_{\rm UV} F + \sigma'_{\rm UV} F'}  , \label{PhotoStabilityM}
\end{equation}
where $\sigma'_{\rm UV}$ and $F'$ is the cross section and the mixing ratio of the interfering species.

Equations (\ref{PhotoStability}, \ref{PhotoStabilityM}) define a critical pressure level (or altitude) for each radiative trace gas subject to photolysis. Above the critical altitude, the gas is readily photodissociated and its mixing ratio decreases with altitude rapidly if no efficient reformation pathway exists; below the critical altitude, photolysis is not important and the gas is likely to be well-mixed. The critical pressure (altitude) is also defined for a certain mixing ratio. If the mixing ratio is larger, the critical pressure is smaller, and the gas is photochemically stable up to a higher altitude. Mean cross sections in dissociating wavelengths and the critical pressure level corresponding to 1 ppm mixing ratio of common spectrally active gases are tabulated in Table \ref{Gas}.

\begin{table}[htdp]
\caption{Mean cross sections in dissociating wavelengths of common atmospheric trace gases and the critical pressures of photolysis. The mean cross sections are weighted by the Solar spectrum and quantum yields in the dissociation wavelengths. The critical pressure is evaluated according to equation (\ref{PhotoStability}) with the conditions of $\mu_0 = 0.5$, $g = 9.8$ m s$^{-1}$, $M_{\rm atm} = 4.65\times10^{-26}$ kg (\ce{N2}-dominated atmosphere), and $F=10^{-6}$ (1 ppm). The additional shielding from other gases and reforming reactions are omitted in the estimate of critical pressures. }
\begin{center}
\begin{tabular}{llll}
\hline
\hline
Name & Dissociation Wavelength & Mean Cross Section & 1-ppm Critical Pressure \\
 & (nm) & (cm$^2$) & (Pa) \\
 \hline
\ce{H2}  &      18 -- 85  &  1.10E-17  &  2.07E+02  \\
\ce{O2}  &       4 -- 245  &  7.18E-21  &  3.17E+05  \\
\ce{H2O}  &       6 --198  &  1.72E-18  &  1.33E+03  \\
\ce{NH3}  &       8 -- 230  &  1.31E-18  &  1.73E+03  \\
\ce{N2O}  &     130 -- 240  &  1.38E-20  &  1.65E+05  \\
\ce{CH4}  &      52 -- 152  &  1.81E-17  &  1.26E+02  \\
\ce{CO2}  &      35 -- 205  &  7.44E-22  &  3.06E+06  \\
\ce{H2S}  &       5 -- 259  &  1.11E-18  &  2.05E+03  \\
\ce{OCS}  &     186 -- 296  &  2.74E-20  &  8.33E+04  \\
\ce{SO2}  &       5 -- 395  &  3.74E-18  &  6.09E+02  \\
\hline
\hline
\end{tabular}
\label{Gas}
\end{center}
\label{default}
\end{table}

\begin{figure}[h]
\begin{center}
 \includegraphics[width=1.0\textwidth]{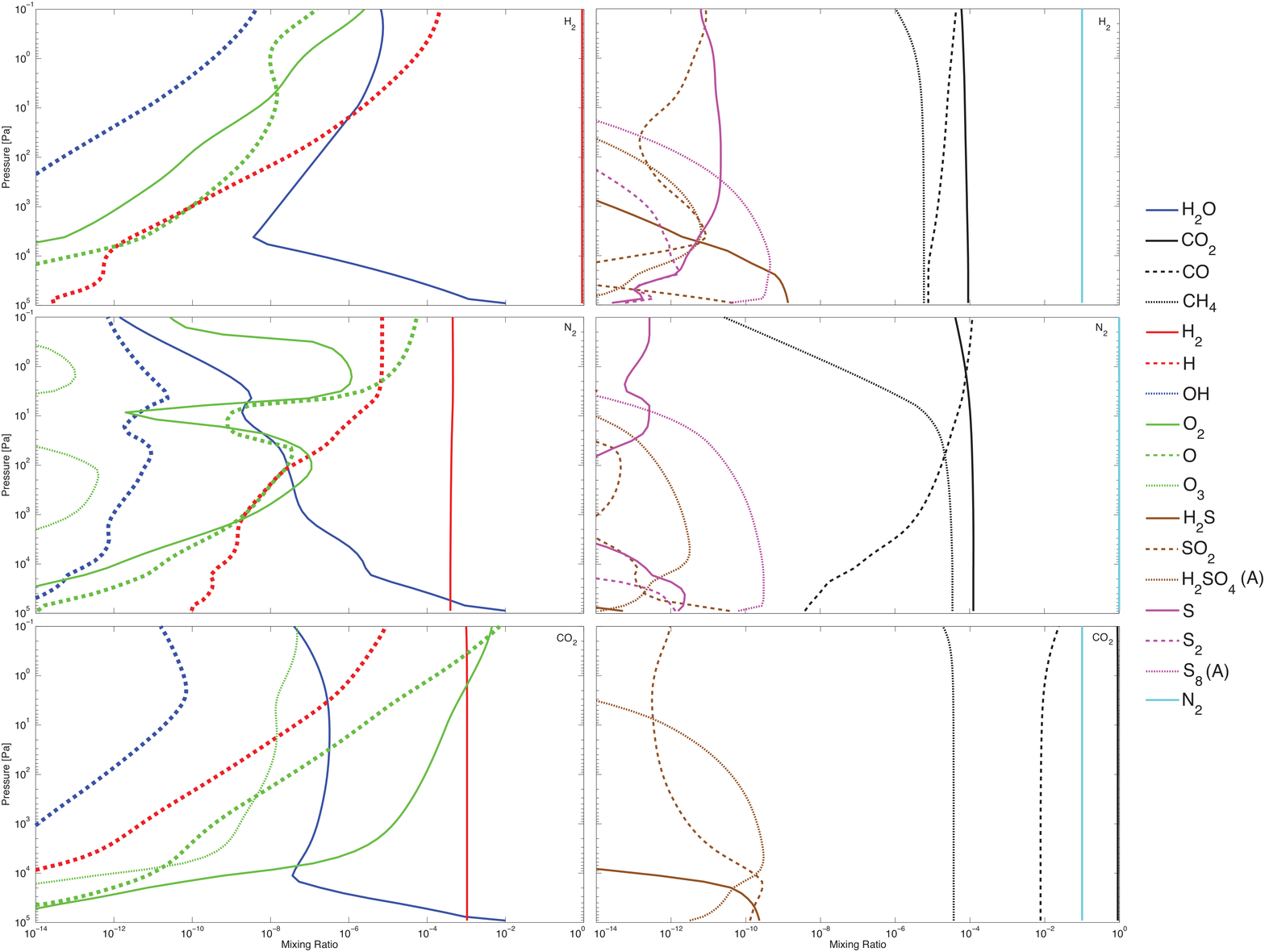}
 \caption{ Compositions of the three scenarios of rocky exoplanet atmospheres tabulated in Table \ref{AtmosScenario}. 
 The left column shows mixing ratios of H and O species, and the right column shows mixing ratios of N, C, and S species.
 From top to bottom, the three panels correspond to the reducing (\ce{H2}-dominated), oxidized (\ce{N2}-dominated), and highly oxidized (\ce{CO2}-dominated) atmospheres. The vertical scales are expressed in pressure, which allows comparison between different scenarios that have very different mean molecular masses, and hence the altitude difference for a given pressure change. We highlight the profiles of three reactive species, \ce{H}, \ce{OH}, and \ce{O} by thick lines.
 }
 \label{Benchmark}
  \end{center}
\end{figure}

Despite simplicity, equations (\ref{PhotoStability}, \ref{PhotoStabilityM}) predict the behavior of trace gases in the upper part of terrestrial exoplanet atmospheres, in agreement with results from the full photochemistry code. For example, 10-ppm methane in the \ce{N2} atmosphere is photodissociated above an altitude of $\sim10$ Pa pressure level, in agreement with Table \ref{Gas}. Using equation (\ref{PhotoStabilityM}) we estimate the 10-ppm critical pressure for methane in the \ce{CO2} atmosphere to be about 0.1 Pa, consistent with results from the full photochemistry code as shown in Figure \ref{Benchmark}. For another example, for \ce{H2S} in the \ce{H2} atmosphere, the near-surface mixing ratio is about $10^{-9}$, and according to Table \ref{Gas}, we find the critical pressure is in the order of $10^5$ Pa, i.e., \ce{H2S} is photodissociated at all altitudes of the \ce{H2} atmosphere. This is again consistent with the full photochemistry model (see the upper panel of Figure \ref{Benchmark}).

Equations (\ref{PhotoStability}, \ref{PhotoStabilityM}) provide an order-of-magnitude assessment on whether or not a vertically well mixed distribution of gas is a good assumption when investigating spectra of super Earths. When the critical pressure defined by equations (\ref{PhotoStability}, \ref{PhotoStabilityM}) is smaller than the pressure level that generates the spectral feature, it is plausible to assume the gas is vertically well-mixed. However, if the critical pressure is larger than the pressure level that generates the spectral feature, cautions should be taken because the gas could be photochemically depleted.
If data suggests an abundance of a certain gas above its critical altitude, an additional mechanism, such as ultraviolet shielding by other gases, or efficient reformation, must be at play.
In summary, equations (\ref{PhotoStability}, \ref{PhotoStabilityM}) provide a simplified approach to take into account the effects of photochemistry for the interpretation of spectral features and for observation planning.

\clearpage

\clearpage

\clearpage

\clearpage

\clearpage

\clearpage

\clearpage

\clearpage

\clearpage

\scriptsize

\begin{table}
\begin{center}
\caption{Photolysis reactions in our photochemistry model. The rates (s$^{-1}$) are computed with the unattenuated Solar spectrum at 1 AU with cross sections and quantum yields at 295 K and 200 K, and divided by 2 for diurnal average.}
\includegraphics[width=1.0\textwidth]{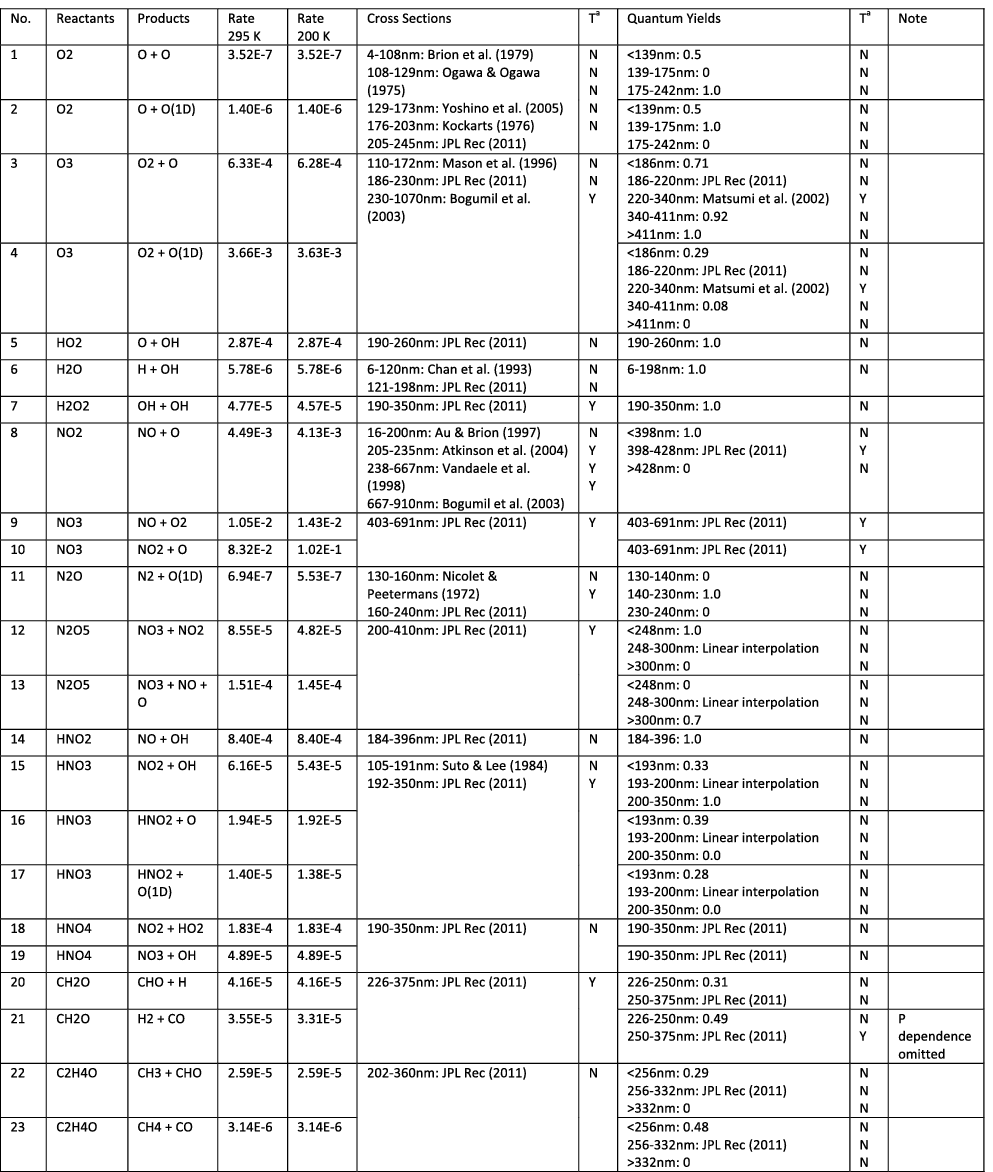}
\label{Photolysis}
\end{center}
\end{table}

\begin{table}
\nonumber
\begin{center}
\caption{Table \ref{Photolysis} continued ...}
\includegraphics[width=1.0\textwidth]{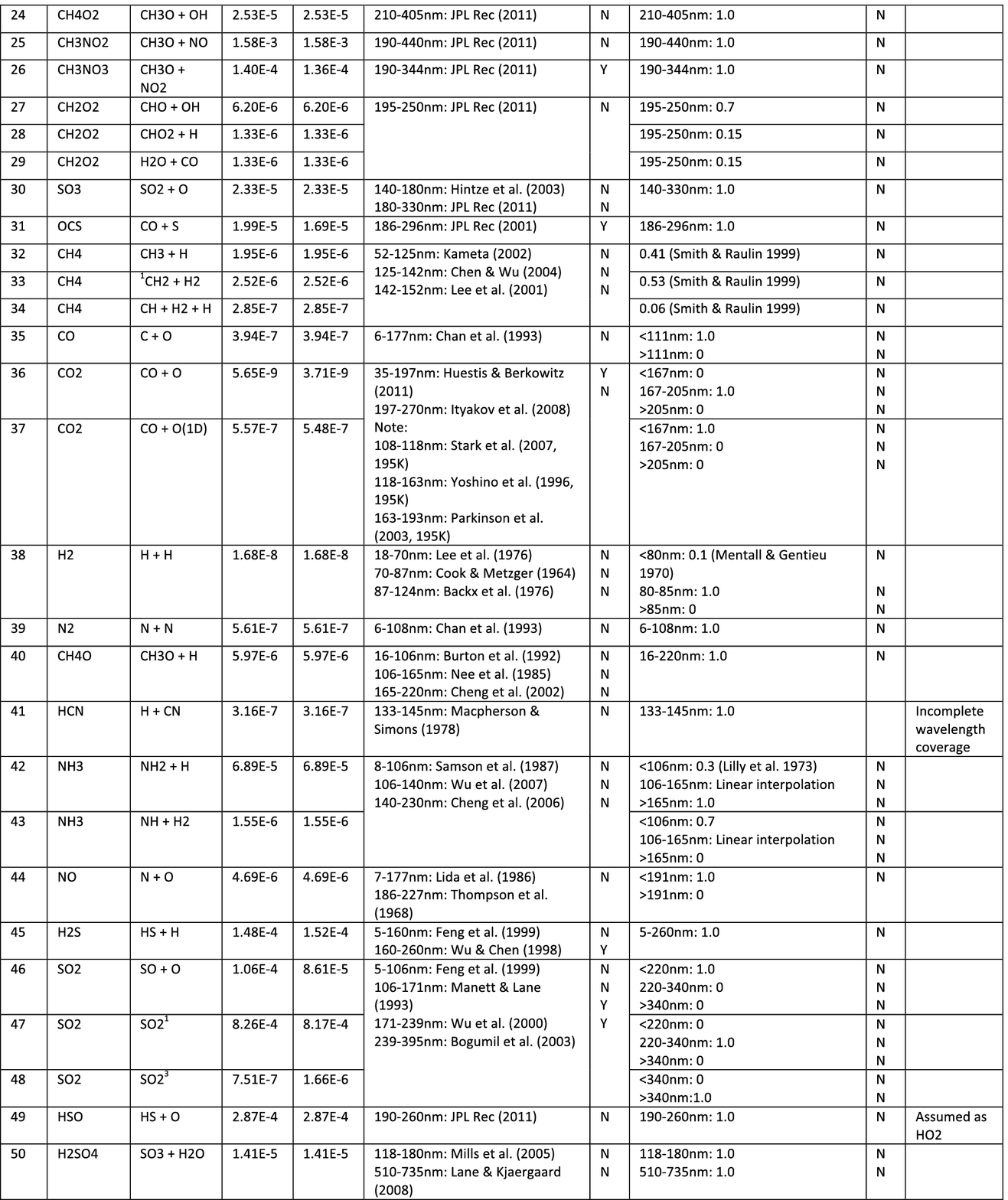}
\nonumber
\end{center}
\end{table}

\begin{table}
\begin{center}
\caption{Table \ref{Photolysis} continued ...}
\includegraphics[width=1.0\textwidth]{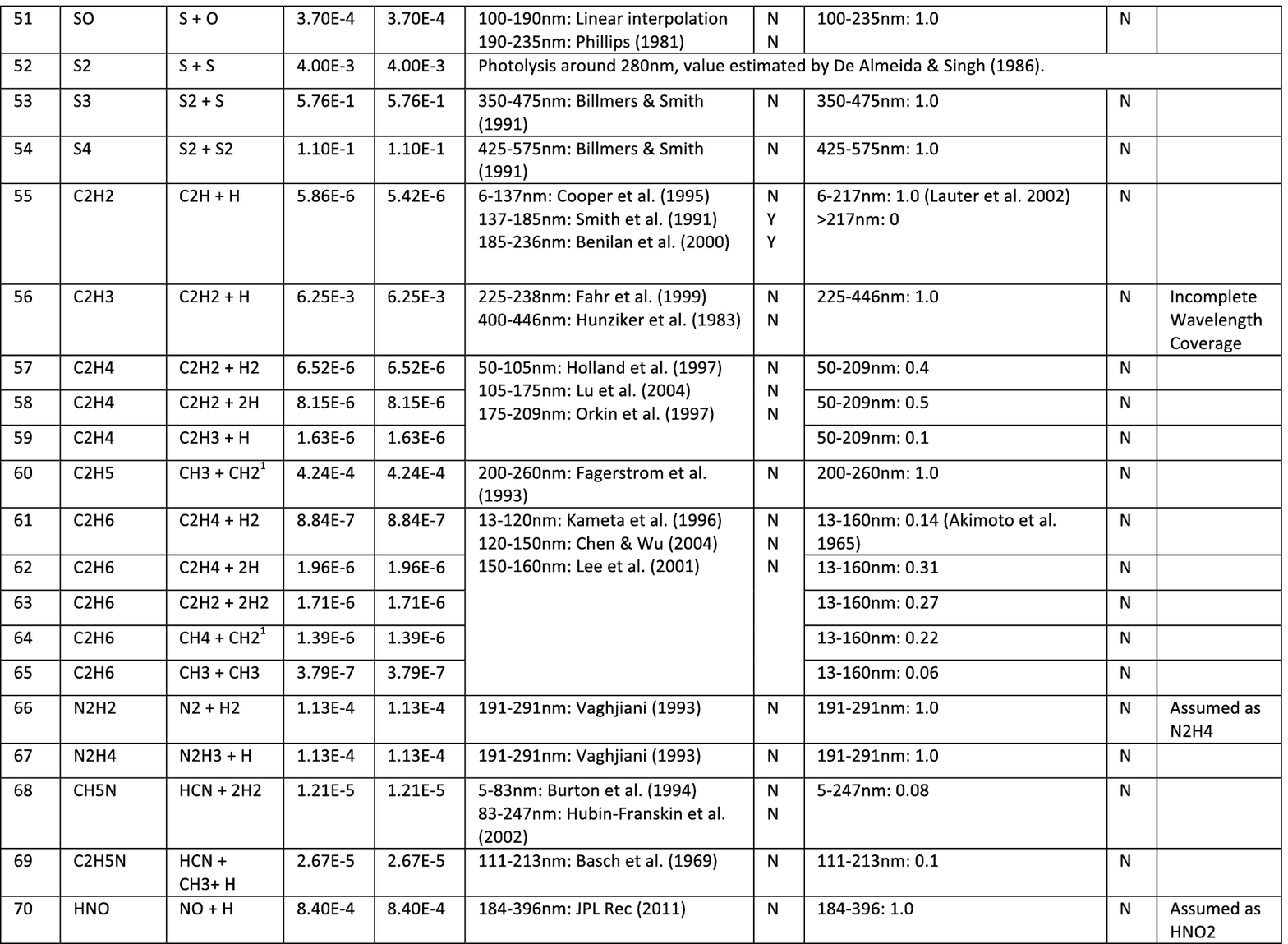}
\tablenotetext{a}{Temperature dependence of cross sections and quantum yields: Y indicates temperature dependence is taken into account for the respective wavelength range.}
\nonumber
\end{center}
\end{table}

\normalsize

\clearpage

\begin{landscape}
\begin{center}

\scriptsize

\begin{longtable}{llllllll}
  \caption{ Reaction rates of bi-molecular reactions (R), ter-molecular reactions (M) and thermo-dissociation reactions (T) in the photochemistry model. In the column ``Temperature" we note the temperature range for which each reaction rate is valid. We annotate "Low Temperature" and "High Temperature" to the reaction rates only valid at temperatures lower than 300 K and higher than 1000 K, respectively. For low-temperature applications, including explorations of Solar-System terrestrial planets and habitable exoplanets, we exclude those reactions marked as ``High Temperature".}
  \label{ReactionRates} \\
   \hline
No. & Reactants & Products & Reaction Rate & Reference & Temperature & Note \\  \hline
  \endfirsthead
     \hline
No. & Reactants & Products & Reaction Rate & Reference & Temperature & Note \\ \hline
  \endhead
  \hline \multicolumn{8}{l}{{Continued on next page ... }} \\
\endfoot
\hline \hline
\endlastfoot

R1 & \ce{ C + CH2 } & \ce{-> CH + CH } & $ 2.69\times 10^{-12} \exp{(-23573.0/T)}  $ & NIST & 1000-4000 & \\
R2 & \ce{ C + CN } & \ce{-> C2 + N } & $ 4.98\times 10^{-10} \exp{(-18041.0/T)}  $ & NIST & 5000-8000 & \\
R3 & \ce{ C + H2 } & \ce{-> CH + H } & $ 6.64\times 10^{-10} \exp{(-11700.0/T)}  $ & NIST & 1520-2540 & \\
R4 & \ce{ C + H2S } & \ce{-> CH + HS } & $ 2.1\times 10^{-10 } $ & NIST & 298 & \\
R5 & \ce{ C + N2 } & \ce{-> CN + N } & $ 8.7\times 10^{-11} \exp{(-22611.0/T)}  $ & NIST & 2000-5000 & \\
R6 & \ce{ C + O2 } & \ce{-> CO + O } & $ 5.1\times 10^{-11} (T/298.0)^{-0.3} $ & NIST & 15-295 & Low Temperature \\
R7 & \ce{ C + OCS } & \ce{-> CO + CS } & $ 1.01\times 10^{-10 } $ & NIST & 298 & \\
R8 & \ce{ C + SO2 } & \ce{-> CO + SO } & $ 6.91\times 10^{-11 } $ & NIST & 298 & \\
\end{longtable}
  \normalsize
\end{center}
\end{landscape}

\end{document}